\documentclass[
twocolumn,
english,
aps,
pra,
longbibliography,
superscriptaddress,
amsmath,
amssymb,
floatfix
]{revtex4-1}
\usepackage[utf8]{inputenc}
\usepackage{microtype}
\usepackage{graphicx}
\usepackage{amsmath}
\usepackage{mathtools}
\usepackage[ruled,lined]{algorithm2e}
\usepackage[pdftex,hyperfigures,pdfpagelabels,colorlinks,citecolor=blue,bookmarks=false,linkcolor=black,urlcolor=blue]{hyperref}
\usepackage{natbib}
\usepackage{braket}
\usepackage{siunitx}
\usepackage{cleveref}
\usepackage{multirow}

\usepackage{graphicx}% Include figure files
\usepackage{dcolumn}% Align table columns on decimal point
\usepackage{bm}% bold math
\usepackage[utf8]{inputenc}
\usepackage{amssymb}
\usepackage{dsfont}
\usepackage{tikz}
\usepackage{qcircuit}
\usepackage{amsfonts,amssymb,amsmath}            % for math symbols.
\usepackage{lmodern,adjustbox}
\usepackage[skins,breakable]{tcolorbox}
\usepackage{xcolor}
\tcbuselibrary{listings}
\tcbuselibrary{breakable}
\usepackage[acronym]{glossaries}
\makeglossaries

\usepackage{graphicx}
\usepackage[caption=false]{subfig}
\renewrobustcmd{\bfseries}{\fontseries{b}\selectfont}
\renewrobustcmd{\boldmath}{}
\newrobustcmd{\B}{\bfseries}

\newacronym{auc}{AUC}{Area Under Curve}
\newacronym{bsm}{BSM}{Beyond-the-Standard Model}

\newacronym{cern}{CERN}{European Organization for Nuclear Research}
\newacronym{hep}{HEP}{High Energy Physics}
\newacronym{hlf}{HLF}{high level feature}
\newacronym{hlt}{HLT}{high level trigger}
\newacronym{hl-lhc}{HL-LHC}{High Luminosity Large Hadron Collider}
\newacronym{lep}{LEP}{Large Electron-Positron Collider}
\newacronym{lhc}{LHC}{Large Hadron Collider}
\newacronym{cms}{CMS}{Compact Muon Solenoid}
\newacronym{mcc}{MCC}{Matthews Correlation Coefficient}
\newacronym{met}{MET}{missing transverse energy}
\newacronym{ml}{ML}{machine learning}

\newacronym{nisq}{NISQ}{Noisy Intermediate-Scale Quantum}

\newacronym{pca}{PCA}{Principal Component Analysis}
\newacronym{pf}{PF}{particle-flow}

\newacronym{roc}{ROC}{Receiver Operator Characteristics}

\newacronym{slac}{SLAC}{Stanford Linear Accelerator Center}
\newacronym{sm}{SM}{Standard Model}
\newacronym{susy}{SUSY}{Supersymmetry}

\newacronym{svm}{SVM}{Support Vector Machine}

\newacronym{qed}{QED}{Quantum Electrodynamics}
\newacronym{qml}{QML}{quantum machine learning}
\newacronym{qnn}{QNN}{quantum neural networks}

\newacronym{qsvm}{QSVM}{Quantum Support Vector Machine}

\newacronym{vbf}{VBF}{vector boson fusion}
\newacronym{vqc}{VQC}{Variational Quantum Classifier}

\newacronym{qgan}{qGAN}{Quantum Generative Adversarial Networks}
\newacronym{gan}{GAN}{Generative Adversarial Networks}
\newacronym{wgan}{WGAN}{Wasserstein Generative Adversarial Networks}
\newacronym{sgd}{SGD}{stochastic gradient descent}

\begin{document}

%\preprint{APS/123-QED}

\title{Quantum Generative Adversarial Networks For \\ Anomaly Detection In High Energy Physics}% Force line breaks with \\

\author{Elie Bermot}
% \email{elie.bermot@polytechnique.edu}
\affiliation{
 IBM Quantum, IBM Research Europe – Zurich, Rueschlikon 8803, Switzerland
}
\affiliation{ETH Zurich, Zurich 8092, Switzerland}
% \altaffiliation[Also at ]{ETH Zurich, Zurich 8092, Switzerland}%Lines break automatically or can be forced with \\
\author{Christa Zoufal}
% \email{ouf@zurich.ibm.com}
\affiliation{
 IBM Quantum, IBM Research Europe – Zurich, Rueschlikon 8803, Switzerland
}

\author{Michele Grossi}
% \email{michele.grossi@cern.ch}
\affiliation{
 European Organization for Nuclear Research (CERN), Geneva 1211, Switzerland
}
\author{Julian Schuhmacher}
% \email{jsc@zurich.ibm.com}
\affiliation{
 IBM Quantum, IBM Research Europe – Zurich, Rueschlikon 8803, Switzerland
}
\author{Francesco Tacchino}
% \email{fta@zurich.ibm.com}
\affiliation{
 IBM Quantum, IBM Research Europe – Zurich, Rueschlikon 8803, Switzerland
}
\author{Sofia Vallecorsa}
% \email{sofia.vallecorsa@cern.ch}
\affiliation{
 European Organization for Nuclear Research (CERN), Geneva 1211, Switzerland
}
\author{Ivano Tavernelli}
% \email{ita@zurich.ibm.com}
\affiliation{
 IBM Quantum, IBM Research Europe – Zurich, Rueschlikon 8803, Switzerland
}

\date{\today}% It is always \today, today,

\begin{abstract}

%The standard model (SM) of particle physics is an accomplishment of decades of theoretical and experimental work. It describes the majority of elementary particles and their interactions: strong, weak, and electromagnetic. However, some of the observed events occurring in a particle accelerator, e.g., the Large Hadron Collider, correspond to anomalous processes, whose underlying physics is not explained by the SM and lot of them are cutted out by conventional model-dependant approaches. 
The standard model (SM) of particle physics represents a theoretical paradigm for the description of the fundamental forces of nature. 
Despite its broad applicability, the SM does not enable the description of all physically possible events.
The detection of events that cannot be described by the SM, which are typically referred to as anomalous, and the related potential discovery of exotic physical phenomena is a non-trivial task.
%experimental and computational 
The challenge becomes even greater with next-generation colliders that will produce even more events with additional levels of complexity.
%--even more so in future experiments since, we expect next generation colliders to be able to access more collision events, including larger number of particles and exhibiting complex correlation. 
The additional data complexity motivates the search for unsupervised anomaly detection methods that do not require prior knowledge about the underlying models.
% \fta{[Can we say something more specific on the computational limitations, if any, that motivate the use of quantum methods? Also, shall we mention the quantum-models-for- quantum-data argument?]} \OUF{Good points. On the former, we need to talk to Michele. The CERN people should probably have a direct argument for generative QML for HEP at hand. For the latter: I'm not sure it fits into the context of the abstract, since we don't have access to any but it would be a good addition to the outlook - if it is not in there already. }
% In fact, until recently most of the high energy physics data analysis relied on model-specific selection processes.  
In this work, we develop such a technique. More explicitly, we employ a quantum generative adversarial network to identify anomalous events. The method learns the background distribution from SM data and, then, determines whether a given event is characteristic for the learned background distribution.
% \fta{[`capacity' and `effective dimension' could sound a bit technical for an abstract. We could add a few words of explanation (i.e., what does model capacity measure?) or some `warnings' (like `so-called effective dimension']}
The proposed quantum-powered anomaly detection strategy is tested on proof-of-principle examples using numerical simulations and IBM Quantum processors. We find that the quantum generative techniques using ten times fewer training data samples can yield comparable accuracy to the classical counterpart for the detection of the Graviton and Higgs particles.
 Additionally, we empirically compute the capacity of the quantum model and observe 
%a potential advantage compared to the classical counterpart.
an improved expressivity compared to its classical counterpart. 
% \OUF{Can we strengthen the statement here? - maybe something with the effective dimension, that we can see an advantage compared to the classical counterpart. AND comparable accuracy with 100 vs. 1000 training data samples. Let's discuss.}

% \begin{description}
% Secondary publications and information retrieval purposes.
% \item[PACS numbers]
% May be entered using the \verb+\pacs{#1}+ command.
% You may use the \texttt{description} environment to structure your abstract;
% use the optional argument of the \verb+\item+ command to give the category of each item. 
% \end{description}
%Use showkeys class option if keyword

\end{abstract}

\maketitle

% \textcolor{red}{[Please double check references. Some duplicates: 27-49, 38-47.]}

\section{\label{sec:level1}Introduction}

%%%%% Why anomaly detection %%%%%%%%
In data science, an anomaly corresponds to a data point that does not fit into the considered data distribution~\cite{anomaly_1, anomaly_2}. Anomaly detection corresponds to the identification of these outliers. Being able to detect these data points is crucial in domains such as network- and cyber-security~\cite{ano_wireless, ano_intrusion}, fraud detection, cancer screening~\cite{ano_cancer}, and many more.

% \fta{[I think this first paragraph could be shortened a bit and linked better with the second one (where we now speak more generically of anomaly detection on `a given dataset').]} 
% Particle physics aims to understand the most elementary components of matter and forces. These elementary particles can be detected in controlled environments with particle accelerators such as the Large Hadron Collider (LHC). 
Another interesting application of anomaly detection is in \gls{hep}. Currently, the \gls{lhc} generates data sets of $O(1)$ MB per sample. Efficiently storing and analyzing the data will become more challenging when the \gls{lhc} will be updated to the \gls{hl-lhc} in 2029, since more collisions will be generated at a time~\cite{hl_lhc}. 
There are many open computational challenges in particle physics concerning data generation, data processing and data analysis~\cite{qml_mc, ref_qml_hep, hep_qml_1, hep_qml_2, hep_qml_3}, e.g., the detection of events that deviate from the \gls{sm}~\cite{ref_article_cern}, the currently most general description of the basic building blocks of the universe. 
% It unifies three of the four fundamental forces that govern the universe: electromagnetism, the strong force, and the weak force. 
While the \gls{sm} is a successful theory that is able to account for many physical phenomena, it is not compatible with, e.g., gravity. In order to fully understand nature, events that
cannot be described by the \gls{sm} need to be investigated in detail. In fact, currently one of the main goals of data
analysis in particle collider experiments is the \gls{ml}-based detection of rare
collisions corresponding to \gls{bsm} particles.
% More specifically, anomaly detection in \gls{hep}  consists in looking for rare collisions that could generate different particles than the ones known from the \gls{sm}. 

%%%%% Why qGAN based anomaly detection %%%%%%%%
% \fta{[The flow of this paragraph can be improved. Also here, I think we can summarize a bit since this is only the introduction, and move the motivation of why generative models can be used for anomaly detection to the theory section. In the intro we need to give the technical context but also get quickly to the motivation and the specific features of our contribution.]} 
% The general goal of anomaly detection is the identification of anomalous events in a given dataset. 
There are two main approaches for \gls{ml}-based anomaly detection, supervised and unsupervised. To train an anomaly detection algorithm in a supervised manner, we require labeled data sets representing the \gls{sm} and \gls{bsm} events. In \gls{hep}, the reference data sets are usually generated through numerical simulations of the \gls{sm} or an \gls{bsm} theory, respectively. Even though this approach has already been applied to detect several rare events in experimental data~\cite{supervised_1,supervised_2,supervised_3,supervised_4}, the main drawback is that it does not allow for the search of particles which are not described by an existing \gls{bsm} theory.
Therefore, recent efforts have been focused on the development of unsupervised approaches~\cite{model_independent,bsm_1,bsm_2,bsm_3,bsm_4,ref_article_anomaly,ano_unsuper_2,ano_unsuper_3}. In the unsupervised settings, a model is trained on unlabeled data, learning its underlying structure with the aim of identifying events that do not conform with the reference data as outliers. If all current knowledge about particle physics is included in the numerical generation of the training data, any outlier identified in a set of measured collisions could then be a sign of new physics. One family of promising unsupervised \gls{ml} approaches for anomaly detection are generative methods~\cite{ref_article_anomaly,ref_article3}, specifically the one based on \gls{gan}s~\cite{GAN_paper,gan_tuto,fanogan2}. Such approaches employ an adversarial training of two classical neural networks, a generator and a discriminator~\cite{efficient_gan_sample_1,efficient_gan_sample_2}. Based on the two networks it is then possible to derive an anomaly score that allows to identify a data point as an anomaly or not.
%{\bf [This paragraph is very confusing, not clear what is ML and what is not...]}}
 % However, training \gls{gan}s requires finding the minimum of non-convex loss functions and could end up in local minima. Therefore, to be robust against training instabilities, both networks can be used to efficiently detect anomalies by designing an anomaly score used to label any data points as an anomaly or not. 

% A recent study~\cite{fewer_data_qml} showed that training \gls{qnn}~\cite{qnn,mangini2021quantum}
% %\textcolor{red}{[Plural seems wrong here]}
% can be done with fewer training data points while giving good generalization errors on unseen data. This implies that there is a potential that \gls{qml} based on \gls{qnn} could in theory be used for \gls{hep} using fewer resources to get efficient representations of the  \gls{sm}.
% We also need to verify whether \gls{qnn} can yield an advantage over classical \gls{ml} methods for \gls{hep} problems. 

In this work, we explore and investigate the feasibility of an unsupervised \gls{qml} approach based on a quantum implementation of a \gls{gan}~\cite{qGAN_1,qGAN_3}, i.e., a \gls{qgan}, that we shall refer to as \textit{Anomaly qGAN} (A-qGAN).
% \gls{qgan}s may have the potential to have an advantage over their classical counterparts to sample probability distributions that are classically intractable~\cite{qgan_advantage_1, advantage_qgan_2, qGAN_2}. 
% which implies that classical data need to be mapped into quantum states using angle encoding~\cite{angle_encoding, angle_encoding_2, angle_encoding_3}.
We demonstrate how a \gls{qgan} may be used to detect \gls{bsm} particles.
% \fta{[We could maybe expand this a bit, and make a more balanced introduction where we highlight better our contributions.]} 
A \gls{qgan} is trained on an embedded SM data set and then used to evaluate an anomaly score by calculating a distance measure between embedded data samples and quantum states generated by the trained \gls{qgan}. Furthermore, we investigate the practical performance of anomaly detection for \gls{bsm} data with quantum simulation and quantum hardware experiments. 
These experiments reveal that an A-qGAN can achieve the same anomaly detection accuracy as its classical counterpart using ten times fewer training data points.
Additionally, we study the expressive power of \gls{qgan}s in the context of anomaly detection w.r.t. a capacity measure -- the so-called \textit{effective dimension}~\cite{effective_dim, eff_dim_1, eff_dim_2, eff_dim_3} -- and find that the quantum models can have advantageous capacity properties.
% Interestingly, using ten times fewer samples than the classical equivalent, our algorithm can yield similar accuracy. 

% The anomaly detection performance gives similar accuracy with less data in the quantum framework compared to the classical algorithm.

% This approach can be interesting because a quantum generator can have a better representation capacity than its classical counterpart. More specifically, a \gls{qgan} approach can cover a bigger model space with less parameters which can be measured with a bigger effective dimension in the parameter space compared to their classical counterpart

This paper is organized as follows: In Section~\ref{sec:theory}, we present both the classical and quantum \gls{gan} models. We then describe the classical and quantum algorithms to perform anomaly detection in Section~\ref{ch:methods}. In Section~\ref{ch:results}, we present the results of empirical anomaly detection experiments with compressed features of an artificial \gls{hep} data set using numerical simulation and actual quantum hardware. Finally, we conclude with a general discussion of anomaly detection with A-qGANs in Section~\ref{ch:conclusions}.

\section{Theory}
\label{sec:theory}
\subsection{Generative Modelling}
\label{ch:generative}

% Introduction generative modeling
The experimental setup considered here corresponds to a black-box setting, where we only have access to the input and output of an experiment.
A powerful algorithm for approximating the underlying process is given by generative learning, which has already been investigated in different scientific domains, including chemistry~\cite{druGAN} and biology~\cite{DNAgan}. 
 In the following, we will focus our study on \gls{gan}s. 
The given task is to learn a representation of the probability distribution underlying a data set $P_{\text{data}}$ using a parameterized ansatz. After a generative model is trained, it can be used to generate new synthetic data that are aligned with the generation process of the training set or to investigate the learned approximation to the model distribution $P_{\text{data}}$. 

\subsubsection{Classical GAN}
\label{sub:gan}
\begin{figure}[htbp!]
\centering
    \includegraphics[width=0.5\textwidth]{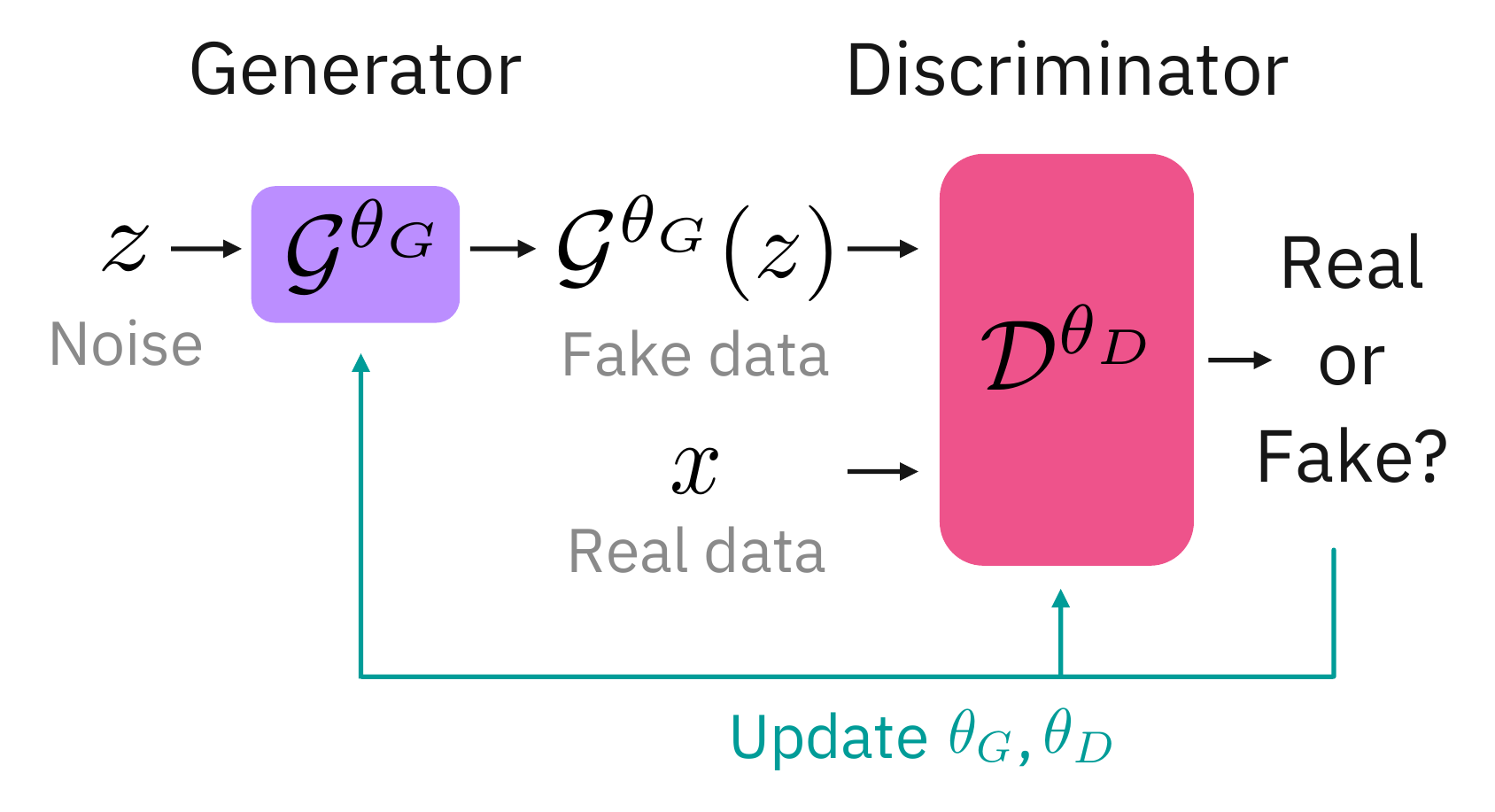}
    \caption{Working principle of the \gls{gan}: a generator $\mathcal{G}^{\theta_{G}}$ samples $k$ data points, and the discriminator $\mathcal{D}^{\theta_{D}}$ receives $k$ data samples at a time from either the real data set and from the generated samples and independently classifies them.}
    \label{GANs}
\end{figure}

\gls{gan}s~\cite{GAN_paper, gan_origin_2} consist of two competing components, the generator $\mathcal{G}^{\theta_{G}}$ and the discriminator $\mathcal{D}^{\theta_{D}}$, each represented by a differentiable, deterministic neural network (see Fig.~\ref{GANs}). The generator $\mathcal{G}^{\theta_{G}}$ aims at generating samples that could be mistaken for being drawn from the data distribution $P_{\text{data}}$ while the discriminator $\mathcal{D}^{\theta_{D}}$ tries to distinguish between the data coming from the generator $\mathcal{G}^{\theta_{G}}$ and samples from the training data set. The generator takes as input a random sample $z$ drawn from a fixed prior distribution $ P_z $ for enabling the generation of multiple outputs.

Multiple loss functions can be used to train this machine learning architecture. 
In this paper, we focus on a non-saturating loss function~\cite{gan_tuto}.  The generator minimizes the following loss function 
\vspace{-2mm}
\begin{equation} \label{eq:cost_classical_generator}
    \text{C}_{G}(\theta_{G}, \theta_{D})=-\frac{1}{2} \mathbb{E}_{z \sim P_{z}} \log \mathcal{D}^{\theta_{D}}(\mathcal{G}^{\theta_{G}}(z)) \,,
\end{equation}
while the discriminator tries to maximize the following loss function
\begin{equation} \label{eq:cost_classical_discriminator}
    \begin{aligned}
    \text{C}_{D}(\theta_{G}, \theta_{D})= & -\frac{1}{2}\mathbb{E}_{z \sim P_{z}} \log(1-\mathcal{D}^{\theta_{D}}(\mathcal{G}^{\theta_{G}}(z)))\\ &-\frac{1}{2}\mathbb{E}_{x \sim P_{\text{data}}}\log \mathcal{D}^{\theta_{D}}(x) \,.
    \end{aligned}
\end{equation}

% In practice,  $\mathcal{D}^{\theta_{D}}(x)$ represents the probability that $x$ came from the data rather than from the generator. 
% Training should improve the performance of both $\mathcal{D}^{\theta_{D}}$ and $\mathcal{G}^{\theta_{G}}$ at their respective tasks simultaneously. 
% The discriminator $\mathcal{D}^{\theta_{D}}$ is trained to maximize the probability of assigning the
% correct label to both training data points and samples from $\mathcal{G}^{\theta_{G}}$.
%The generator $\mathcal{G}^{\theta_{G}}$ is trained at the same time to minimize the difference between real and fake labels
%\log(1 - \mathcal{D}^{\theta_{D}}(\mathcal{G}^{\theta_{G}}(z)))$. It corresponds to a min-max optimization problem with the value function $V(\mathcal{G}^{\theta_{G}}, \mathcal{D}^{\theta_{D}})$: 
% \begin{multline} 
%     \min_{\mathcal{G}^{\theta_{G}}} \max_{\mathcal{D}^{\theta_{D}}} V(\mathcal{D}^{\theta_{D}}, \mathcal{G}^{\theta_{G}}) =  \mathbb{E}_{x\sim p_{data}}[\log(\mathcal{D}^{\theta_{D}}(x))] \\
%      +\mathbb{E}_{z\sim p_{z}}[\log(1-\mathcal{D}^{\theta_{D}}(\mathcal{G}^{\theta_{G}}(z)))].
% \end{multline}
    
% This problem can be reduced to the optimization of two non-saturating loss functions~\cite{gan_tuto} for the generator and for the discriminator respectively. The discriminator's loss functions reads: 

% Several techniques exist to specify the cost function of the generator~\cite{GAN_paper}. We use the heuristically generator loss function: 

For the optimization of the parameters $\theta_G$ and $\theta_D$,  standard optimizers can be employed, such as stochastic gradient descent~\cite{sgd}, ADAM~\cite{adam} or AMSGRAD~\cite{amsgrad}, with alternating updates for generator and discriminator parameters.

\subsubsection{Quantum GAN}
\label{sub:qgan}
% In this section, we introduce the quantum framework of the classical GAN that we are using for anomaly detection. It focuses on the training of a fully quantum GAN with a quantum data set, a quantum generator and a quantum discriminator. It is an advantageous structure: the quantum generator can be used to learn an underlying quantum process, and a quantum discriminator only needs to measure a single qubit. This discriminator approach helps reducing measurement errors and avoids using global objective functions which could lead in the vanishing gradient problem, so called, barren plateaus~\cite{bp_1, bp_2}. 

\begin{figure}[htpb]
    \centering 
    \includegraphics[width=0.5\textwidth, trim={0 0cm 0 0cm}, clip]{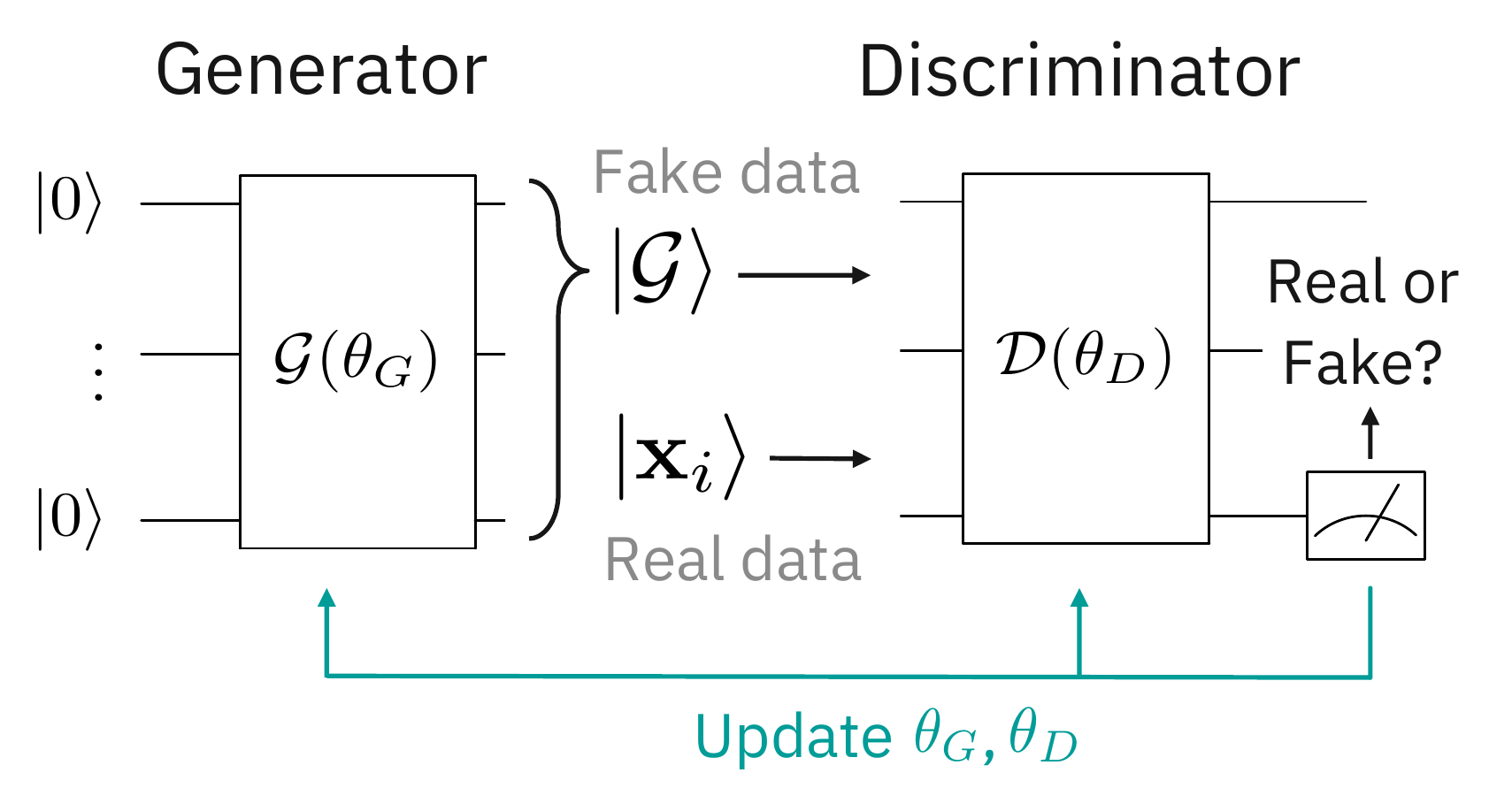}
    \caption{Illustration of a \gls{qgan} model for learning the distribution underlying a training data set $\{ \ket{\mathbf{x}_{1}}, \ldots, \ket{\mathbf{x}_{M}} \}$. At each training step, the generator generates a state $\ket{\mathcal{G}}$ and the discriminator labels the generated state  and a batch of the embedded data points  $\ket{\mathbf{x}_{i}}=\mathcal{R}_{i}\ket{0}$ as real or fake depending on the $Z$-measurement of the last qubit.  }
    \label{circuit_pure}
\end{figure}

Several approaches are possible to translate the \gls{gan} framework into a quantum machine learning context~\cite{qGAN_1, qGAN_3, qGAN_2, qml_mc}. 
In this work, we realize the generator and the discriminator with a parameterized quantum circuit, as is illustrated in Fig.~\ref{circuit_pure}. 
Both, the generator $\mathcal{G}$ and the discriminator $\mathcal{D}$, are defined on $n$-qubit registers with parameterized unitaries $\mathcal{G}(\theta_G)$ and $\mathcal{D}(\theta_D)$. The generator aims at generating a quantum state that represents the distribution underlying the training data set. The discriminator classifies data points as real/generated depending on a single-qubit measurement on the discriminator's qubit register.  

To enable a compatibility of this quantum model with classical training data, we need to map each classical training data point $\{\mathbf{x}_{1}, \ldots, \mathbf{x}_{M} \}$ to a quantum state $\{\ket{\mathbf{x}_{1}}, \ldots, \ket{\mathbf{x}_{M}} \}$. It should be noted that the chosen mapping needs to be efficient.
%The classical training data need to be mapped individually in an efficient way to the respective quantum states. 
The generator circuit outputs a state $\ket{\mathcal{G}}=\mathcal{G}(\theta_G)\ket{0}^{\otimes n}$. Generating samples from our model requires to take measurements from $\ket{\mathcal{G}}$ which are then mapped to the feature space  $\{\mathbf{x}_{1}, \ldots, \mathbf{x}_{M} \}$.  The discriminator takes as input an $n$-qubit quantum state and labels it as real or generated based on the measurement of the Pauli $Z$ observable on the last qubit. If the resulting measurement corresponds to the $-1$ eigenvalue, the input data is classified as real and otherwise as generated. 

To simplify the notation, we introduce the following helper functions:
% \begin{widetext}
\begin{align}
    \text{C}^{\text{generated}}(\theta_{G}, \theta_{D}) = \frac{1}{2} - \frac{1}{2} \braket{ \mathcal{G} | \mathcal{D}^{\dagger}(\theta_D) \, \hat{O} \, \mathcal{D}(\theta_D)| \mathcal{G}}, \label{eq:helper_G} \\
    \text{C}^{\text{data}}(\theta_D) = \frac{1}{2} - \frac{1}{2M}\sum_{i=1}^{M} \braket{\mathbf{x}_{i} |\mathcal{D}^{\dagger}(\theta_D) \, \hat{O} \, \mathcal{D}(\theta_D)|\mathbf{x}_{i}}, \label{eq:helper_R} 
\end{align}
% \end{widetext}
where $\hat{O} = \left( \mathds{1} \otimes \mathds{1} \otimes \ldots \otimes Z \right)$ and $M$ denotes the number of data samples.
%with $\ket{\mathcal{G}}$ denoting the generator state, each $\ket{\mathbf{x}_i}$ denoting each training data point and $M$ denoting the total number of data points. 
% The respective functions determine the probability $\mathbb{P}$ of the discriminator measurement corresponding to $-1$. \\
Both functions determine the probability of the discriminator labelling a state as real. \\

%Firstly, the generator $\mathcal{G}(\theta_G)$ intends to generate a state $\ket{\mathcal{G}}$ that reproduces the desired quantum data set $\{\ket{\mathbf{x}_{1}}, \ldots, \ket{\mathbf{x}_{M}} \}$. 
The loss function for the generator is then given by
\begin{equation}\label{loss_generator}
    \text{C}_{G}(\theta_{G}, \theta_{D}) =-\text{C}^{\text{generated}}(\theta_{G}, \theta_{D}).  
\end{equation}
The probability for the generated state $\ket{\mathcal{G}}$ to be labeled as real by the discriminator is maximized by minimizing $\text{C}_{G}(\theta_{G}, \theta_{D})$ with respect to $\theta_{G}$.
%with $\text{C}_{D}^{\mathcal{G}}$ given by Eq.~\eqref{eq:helper_G}. The loss function aims at maximizing the probability for the generated state $\ket{\mathcal{G}}$ being labeled as real by the discriminator.
The discriminator $\mathcal{D}(\theta_D)$, on the other hand, tries to discriminate between the generator output and training data by  maximizing the following cost function with respect to $\theta_D$:
%The loss function of the discriminator reads: 
\begin{equation} \label{loss_discriminator}
    \text{C}_{D}(\theta_G, \theta_D)=   C^{\text{generated}}(\theta_G, \theta_D)-\text{C}^{\text{data}}(\theta_D).
\end{equation}
%with $\text{C}_{D}^{\mathcal{R}}$ given by Eq.~\eqref{eq:helper_R}. In other words, the goal is to maximize the probability that the quantum data set $\{\ket{\mathbf{x}_{1}}, \ldots, \ket{\mathbf{x}_{M}} \}$ is labeled as real by the discriminator.
During the training the discriminator receives the quantum generator output, i.e. a single quantum state representing the full spectrum of samples, and a batch of $M$ individual samples $\{\ket{\mathbf{x}_i}\}$ from the training data. This may induce a bias in the training that should be monitored carefully and potentially counteracted with multiple optimization steps for the discriminator before performing a single optimization step for the generator.

The loss functions in Eqs.~\eqref{loss_generator} and \eqref{loss_discriminator} are optimized in an alternating fashion with, e.g., the ADAM optimizer, by computing analytic quantum gradients~\cite{analytic_gradient} using a parameter shift rule~\cite{parameter_shift_1, parameter_shift_2}. 
To ensure trainability, i.e. avoid exponentially vanishing gradients, the ansatz should be chosen sufficiently shallow and with a suitable entanglement structure~\cite{bp_1, Wiebe2020Barren}. Furthermore, the method can be expected to be stable against cost-function dependent barren plateaus~\cite{bp_2} since only a single qubit measurement is applied in the discriminator.
% \fta{Don't we have also the fidelity somewhere? That would not fall in this category.} \elie{The fidelity is only considered for the anomaly score, not for the training.}

\subsection{Anomaly Detection}
\label{ch:A-GAN}
% \fta{[This paragraph could be very good for the intro ;)]} 
\subsubsection{Classical benchmark}
\label{sub:classical}

% \OUF{First introduce the AnoGAN concept. We've spoken about it before in the intro but the setting should be described in more detail here.
% } 
% In the ideal case, the trained discriminator could directly enable anomaly detection by classifying anomalous data as fake data. In practice, however, it is unlikely that the discriminator is trained perfectly. More explicitly, the adversarial training of both the generator and the discriminator typically induces unstable learning behavior resulting in the possibility of either the generator or the discriminator being stronger. Thus, to avoid this difficulty, practical anomaly detection rather relies on designing a score using both the generator and discriminator to determine whether a particular event is uncharacteristic of a given background distribution. 

The \gls{gan} framework introduced in Ref.~\cite{ref_article3} uses the adversarial nature of the two networks to flag anomalous data points. In the training, the generator learns the structure of the non-anomalous data samples. Then, an anomaly score is calculated on the testing data set and/or on new data points to determine whether a new sample aligns with the data distribution. This anomaly score is obtained by optimizing an anomaly loss function based on a combination of the  discriminator and generator output. This realizes a combination of all possible information available to the system and can help to compensate for potential instabilities in the generator/discriminator training.

Given a data point $\mathbf{x}$, the minimization of the anomaly loss function aims to find an input noise $z_{\text{opt}}$ corresponding to a generated event $\mathcal{G}(z_{\text{opt}})$ as similar as possible to $\mathbf{x}$ and located on the learned manifold of the \gls{gan}. The corresponding anomaly loss function reads
\begin{align}\label{anoscore_classical}
    \mathcal{S}_{\mathcal{C}}(\mathbf{x};\alpha)&=\min_{z} \mathcal{L}_{\mathcal{C}}(z) \nonumber \\&=\min_{z} \, (1-\alpha)\lVert \mathbf{x} - \mathcal{G}(z) \rVert + \alpha \lVert \mathcal{D}(\mathbf{x}) - \mathcal{D}(\mathcal{G}(z)) \rVert.
\end{align}
The first term computes the similarity between the generated event $\mathcal{G}(z)$ and the data point $\mathbf{x}$ and the second term measures the similarity of the discriminator outputs $\mathcal{D}(\mathcal{G}(z))$ and $\mathcal{D}(\mathbf{x})$.
%feature similarity between the generated event $\mathcal{G}(z)$ and the data point $\mathbf{x}$ by measuring the distance of the output of the discriminator $\mathcal{D}$.
%  For a perfect discriminator,  we have $\mathcal{D}(\mathbf{x}) = \mathcal{D}(\mathcal{G}(z_{\text{opt}}))$ for a perfect discriminator. 
Furthermore, the parameter $\alpha$ is used to weigh the importance of both terms.

Given $\alpha$ and a well-chosen threshold $\delta > 0$, a data point $\mathbf{x}$ is identified as an anomaly if
\begin{align} \label{ano_threshold}
    \mathcal{S}_{\mathcal{C}} (\mathbf{x};\alpha) >\delta.    
\end{align}
% In practice, we tune the threshold $\delta$ according to the given training data as the value that maximizes a set of classification metrics.
In practice, the threshold $\delta$ is chosen such that it realizes a representative differentiation between non-anomalous and anomalous data. In our case, we may employ training data corresponding to \gls{sm} and \gls{bsm} to find a reasonable baseline for this parameter.

% Different variations of this approach are possible, i.e., a Wasserstein \gls{gan} with gradient penalty can be used instead of a standard \gls{gan} as in~\cite{fanogan}. However, this approach requires an optimization for each testing data point to recover the optimal $z_{\text{opt}}$ using stochastic gradient descent for example. This procedure is computationally expensive as every gradient computation requires backpropagation through the generator network.  To speed up the anomaly detection scheme, the generator is usually coupled with a variational autoencoder~\cite{survey_gan_ano}  or also with a bi-directional \gls{gan}, which is then called the BiGAN model~\cite{ efficient_gan, efficient_gan_2}.  \\

\subsubsection{Quantum approach}
\label{sub:quantum}
Next, we introduce our A-qGAN algorithm for the detection of anomalies.
% The training of a qGAN can also be subject to 
We consider a \gls{qgan} architecture with a quantum generator and a quantum discriminator as described in Section~\ref{sub:qgan}. First, each classical data point $\mathbf{x}_{j}$ is mapped to a quantum state $\ket{\mathbf{x}_{j}}$. During the training, the \gls{qgan} learns a representation for the distribution underlying the generation process of the classical $\{\mathbf{x}_{j}\}_{j=1}^{M}$. The anomaly score is then evaluated for data points $\ket{\mathbf{x}}$ of the test data set. The loss for the anomaly score is defined as: \\
% \begin{align}
\begin{equation}\label{anoscore_quantum}
\mathcal{S}_{\mathcal{Q}}(\mathbf{x};\alpha)=(1-\alpha) \lvert \braket{\mathbf{x} |  \mathcal{G}} \rvert^{2} + \alpha \displaystyle \frac{1 + \braket{Z}_{\mathcal{D}\ket{\mathbf{x}}} \braket{Z}_{\mathcal{D}\ket{\mathcal{G}}}}{2},
\end{equation}
% \begin{align}\label{anoscore_quantum}
    % \mathcal{S}_{\mathcal{Q}}(\mathbf{x};\alpha)=&\sum\limits_{j=1}^M\Big((1-\alpha) \lvert \braket{\mathbf{x}_j |  \mathcal{G}} \rvert^{2} \nonumber \\
    % &+ \alpha \displaystyle \frac{1 + \braket{Z}_{\mathcal{D}\ket{\mathbf{x}_j}} \braket{Z}_{\mathcal{D}\ket{\mathcal{G}}}}{2}\big),
% \end{align}
with $\braket{Z}_{\ket{u}}=\braket{u | \mathds{1} \otimes \mathds{1} \otimes \ldots \otimes Z |u}$. The first term is a \textit{residual score}. It measures the fidelity between the embedded input point $\ket{\mathbf{x}_j}$ and the trained generator state $\ket{\mathcal{G}}$. 
% For a non-anomalous data point $\mathbf{x}$, a perfect generator yields $\lvert \braket{\mathbf{x} | \mathcal{G}}\rvert^2=1$. 
It should be noted that the fidelity measure can suffer from exponential concentration effects~\cite{fidelity_scable,tacchinoIEEE2021}. Finding scalable distance measures remain an open question for future research. 
The second term is a \textit{discriminator score} based on the classification of the embedded input point $\ket{\mathbf{x}_j}$ and the trained generator state $\ket{\mathcal{G}}$. %It is computed as follows: we load the state $\ket{\mathbf{x}_j}$ into the discriminator and measure the last qubit of $\mathcal{D}$.
Finally, the parameter $\alpha$  weighs the importance of both circuits for the anomaly score. Equivalently to the classical case, an event is labelled as anomalous if \begin{align} \label{qano_threshold}
    \mathcal{S}_{\mathcal{Q}} (\mathbf{x};\alpha) >\delta,   
\end{align}
for reasonably chosen $\delta$ and $\alpha$.
The quantum anomaly score differs from the classical score in the sense that it does not require an additional optimization over the random prior $z$ due to the intrinsic stochasticity of quantum measurements.

\section{Methods}
\label{ch:methods}

\subsection{High Energy Physics Data Set}
\label{sub:dataset}
In this work, we are focusing on an artificial data set generated according to \gls{hep} experiments. 
The training data set consists of a weighted mixture of different \gls{sm} processes typically observed at $13$ TeV with weights given by the production cross section of the corresponding processes~\cite{top}. This accounts for the most representative processes in the \gls{sm}. The anomalous data sets were obtained through the simulation of \gls{bsm} processes obtained with the PYTHIA8 Monte Carlo simulator~\cite{monte_carlo, monte_carlo_2, monte_carlo_graviton}. In this work, we consider the Higgs boson and the Graviton as anomalies. %data sets are considered as anomalies in this work.
The \gls{sm} training data set contains 3'450'279 different events and the testing data set contains 3'450'277 events. 
Additionally to the \gls{sm} data set, we also work with a Higgs boson data set~\cite{dataset_higgs} containing 139'991 events and a Graviton data set~\cite{dataset_graviton} containing 6'910 events. 
In all data sets, an event is characterized by a list of 23 high-level features~\cite{hlf}.
For the quantum approach, we randomly select 100 \gls{sm} events to build a training set and 100 events for each the testing \gls{sm} data set, the Graviton data set and the Higgs data set. For the classical benchmark, we randomly select 10 times more, i.e., 1'000 events from each data set including the 100 events used for the quantum approach. 

We apply classical pre-processing steps: we efficiently extract a compressed representation of the data sets with the \gls{pca} method~\cite{pca} and normalize the training data set to be in $[-\frac{\pi}{2} , \frac{\pi}{2}]$. This is necessary due to the periodicity of Pauli rotations that builds the foundation of the data encoding in our ansatz.  

The training quantum data set is obtained by mapping each data point $\mathbf{x} \in [-\frac{\pi}{2} , \frac{\pi}{2}]^{n}$ to an $n$-qubit quantum state $\ket{\mathbf{x}}$ obtained with the angle encoding~\cite{angle_encoding, angle_encoding_2, angle_encoding_3}:
\begin{equation}
       \ket{\mathbf{x}} = \bigotimes_{i=1}^{n} R_{y}(x_i) \ket{0}_{i},
\end{equation}
where $x_{i}$ corresponds to the feature $i$ of the data point $\mathbf{x}$ and $\ket{0}_{i}$ to the ground state of qubit $i$.

\subsection{Quantum implementation}
\label{sub:quantum_implementation}
\begin{figure*}[htpb!]
\centering 
\subfloat[Ansatz for the generator \label{sfig:generator_ansatz}]{
\scalebox{0.75}{
\Qcircuit @C=0.8em @R=1.2em {
 & &  & &  \mbox{repeat $k_{G}$ times } &&&&&&\\
 \lstick{\ket{0}} & \gate{H} & \qw & \gate{R_{Y} (\theta^{0,j }_{G})}  & \ctrl{1} & \qw  & \qw  & \qw  & \ctrl{3}  &  \qw & \qw &\qw & \qw\\
 & \vdots &  & \vdots   &  &  &   &   & && \\
 \lstick{\ket{0}} & \gate{H} &\qw & \gate{R_{Y} (\theta^{i,j }_{G})} & \ctrl{-1} & \qw  & \ctrl{1} & \qw  & \qw & \qw & \qw&\qw &\qw\\
   & \vdots &  & \vdots   &  &  &   &  & &&  \\
 \lstick{\ket{0}}& \gate{H} & \qw&  \gate{R_{Y} (\theta^{n-1,j }_{G})} & \qw &\qw  & \ctrl{-1} & \qw  & \ctrl{-3} & \qw & \qw  \gategroup{2}{4}{6}{11}{1em}{--} &\qw & \qw
 }}}
 \hspace{10mm}
 \subfloat[Ansatz for the discriminator \label{sfig:discriminator_ansatz}]{\scalebox{0.8}{
\Qcircuit @C=0.5em @R=1.2em {
 & &  & &  \mbox{repeat $k_{D}$ times } &&&&&&\\
 \lstick{\ket{0}} & \gate{H} & \qw & \gate{R_{Z} (\theta^{0,j }_{D})}  &  \gate{R_{Y} (\theta^{0,j+1 }_{D})} &\gate{R_{Z} (\theta^{0,j+2 }_{D})}&  \ctrl{1} & \qw  & \qw  & \qw    &  \qw & \qw & \qw &\ctrl{4}& \qw &\qw &\qw&\qw &\qw & \qw   \\
 & \vdots &  & \vdots   &  &  &   &   & &  &&&&&&\\
 \lstick{\ket{0}} & \gate{H} &\qw & \gate{R_{Z} (\theta^{i,j }_{D})}  &  \gate{R_{Y} (\theta^{i,j+1 }_{D})} &\gate{R_{Z} (\theta^{i,j+2 }_{D})} & \ctrl{-1} & \qw  & \ctrl{1} & \qw  & \qw & \qw & \qw &\qw&\qw&\ctrl{2}&\qw & \qw &\qw & \qw  \\
   & \vdots &  & \vdots   &  &  &   &  & &  &&& &&&&&\\
 \lstick{\ket{0}}& \gate{H} & \qw&  \gate{R_{Z} (\theta^{n-1,j }_{D})}  &  \gate{R_{Y} (\theta^{n-1,j+1 }_{D})} &\gate{R_{Z} (\theta^{n-1,j+2 }_{D})}  & \qw &\qw  & \ctrl{-1} & \qw  & \qw & \qw  \gategroup{2}{4}{6}{11}{1em}{--}& \qw &\targ{}&\qw&\targ{}&\gate{R_{X}(\theta_{D}^{n, 4})} & \gate{R_{Y} (\theta^{n,5 }_{D})}  & \gate{R_{Z} (\theta^{n,6}_{D})} & \qw 
 }}}
\caption{Ansatz for the quantum architecture. The quantum generator (a) consists of $R_{y}$ rotation gates and $CZ$ gates with variational parameters $\theta^{i, j}_{G}$ which corresponds to the parameter of the $R_{Y}$ rotation acting on the $i$ qubit on the $j$ layer. The quantum discriminator (b) consists of $R_{X}$, $R_{Y}$ and $R_{Z}$ rotation gates, $CZ$ and $CNOT$ gates with variational parameters $\theta^{i, j}_{D}$.}

\label{quantum_ansatz}
\end{figure*}
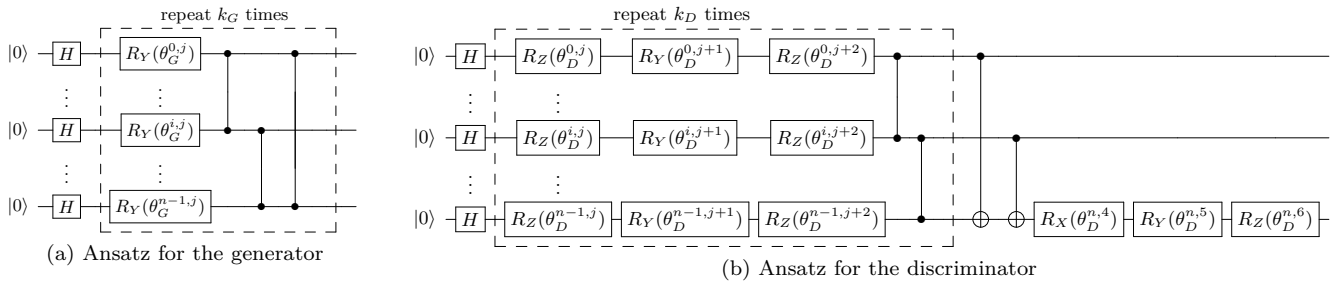

We implement the quantum anomaly detection algorithm in Qiskit~\cite{qiskit}. The quantum generator circuit $\mathcal{G}(\theta_{G})$ consists of an initialization layer and $k_{G}$ alternating layers of parameterized Pauli-$Y$ single-qubit rotations and blocks of  controlled-$Z$ gates (see Fig.~\ref{sfig:generator_ansatz})~\cite{pqc, hea}. The initialization layer consists of Hadamard gates $H$~\cite{qc} on all qubits. In total, the circuit has $(k_{G}+1)n$ parameterized Pauli-$Y$ rotations and $k_{G}n$ controlled-$Z$ gates.

The quantum discriminator is also a parameterized quantum circuit. We use an ansatz containing approximately as many parameters as the generator. The variational form of the discriminator consists of several parts: first, we apply a layer of Hadamard gates on all qubits followed by $k_{D}$ alternating layers of parameterized single-qubit rotations, Pauli-$Y$($R_{Y}$) and Pauli-$Z$($R_{Z}$) rotations on all qubits followed by controlled-$Z$ gates ($CZ$). Then, a series of controlled-$X$ ($CNOT$)~\cite{qc} is applied with the target on the last qubit. Finally, $R_{X}$, $R_{Y}$ and $R_{Z}$ single-qubit rotations are applied on the last qubit. In total, the circuit has $3(k_{D}n+1)$ parameterized single-qubit rotations.  The quantum circuit of the discriminator is shown in Fig.~\ref{sfig:discriminator_ansatz}.

In practice, at each optimization step, the training data is shuffled and split into batches of size 10. The discriminator is trained with samples from the batch while the generator is trained to output a single quantum state. We improve the stability of the algorithm with more optimization iterations for the discriminator than for the generator. The optimizers and respective hyperparameters which are used in the experiments are given in Section~\ref{ch:results}.

\subsection{Classification scores}
\label{sub:classification}

We measure the performance of this anomaly detection problem with different metrics, the \gls{roc}~\cite{auc_1, auc_2}, accuracy, F1 score, and precision, which are defined as: 

\begin{align}
   \text{Accuracy} &= \frac{\text{True Anomalies}}{\text{Total number of events}},\\
%   \text{Recall}&=\frac{\text{True Anomalies}}{\text{True Anomalies} + \text{False Non-Anomalies }}, \\
   \text{Precision}&=\frac{\text{True Anomalies}}{\text{True Anomalies} + \text{False Anomalies}},\\
   \text{and} \nonumber \\
   \text{F1}&=\frac{2}{\text{Recall}^{-1} + \text{Precision}^{-1}} \\
   \text{with Recall}& = \frac{\text{True Anomalies}}{\text{True Anomalies}+\text{False Non-Anomalies}} \nonumber.
\end{align}
An \gls{roc} curve corresponds to the plot of the true positive rate versus the false positive rate. It essentially measures the performance of a binary classifier as its cutoff threshold is varied. In  practice,  the  area  under  the  \gls{roc}  curve,  the  \gls{auc}, represents the degree of separability of a classifier. The higher the \gls{auc} the better the model is at predicting the correct class of its input. 
\subsection{Anomaly detection algorithm}
\label{sub:ano_algorithm}
% In both the quantum and the classical frameworks, we detect anomalies using the same approach.
The anomaly score is computed for each sample in the \gls{sm}, Graviton and Higgs data sets. The mean of the anomaly scores for the data sets define thresholds which in turn help to identify whether a test sample is an anomaly or not. 
We apply a grid search over the $\alpha$ parameter, to find the one which results in the best AUC on the test data set.
The anomaly detection procedure is explained in Algorithm~\ref{alg:anomaly_algorithm}. 

\begin{algorithm}
\caption{Anomaly detection algorithm}\label{alg:anomaly_algorithm}
% \begin{algorithmic}[1]
  \SetAlgoLined
 \SetKwData{Left}{left}\SetKwData{This}{this}\SetKwData{Up}{up}
\SetKwFunction{Union}{Union}\SetKwFunction{FindCompress}{FindCompress}
\SetKwInOut{Input}{input}\SetKwInOut{Output}{output}
\KwData{Training data set 
$\mathbb{D}=\{\mathbf{x}_{1}, 
\ldots, \mathbf{x}_{M} \}$ (or
$\{\ket{\mathbf{x}_{1}}, \ldots, \ket{\mathbf{x}_{M}}\}$), 
testing data set $\mathcal{T}$, generator optimizer $\text{Opt}_{G}$, discriminator optimizer $\text{Opt}_{D}$.}
\For{number of training steps}
    {Train generative model $(\mathcal{G}, \mathcal{D})$ based on Section.~\ref{sub:gan} or~\ref{sub:qgan} with $\text{Opt}_{G}$ and $\text{Opt}_{D}$.}
\For{$\alpha \in [0 , 1]$} {
    \For{$\mathbf{x} \in \mathbb{D}\cup \mathcal{T}$}{
    \If{classical}{
        \scalebox{0.83}{ $\mathcal{S}_{\mathcal{C}}(\mathbf{x}) \gets \min_{z} [(1- \alpha) \lVert \mathbf{x}-\mathcal{G}(z) \rVert + \alpha \lVert \mathcal{D}(\mathbf{x})-\mathcal{D}(\mathcal{G}(z)) \rVert $}; }
    
    \If{quantum}{
        \scalebox{0.9} {$ \mathcal{S}_{\mathcal{Q}}(\mathbf{x}) \gets (1-\alpha) \lvert \braket{\mathbf{x} |  \mathcal{G}} \rvert^{2} + \alpha \displaystyle \frac{1 + \braket{Z}_{\mathcal{D}\ket{\mathbf{x}}} \braket{Z}_{\mathcal{D}\ket{\mathcal{G}}}}{2}$};}
    
}
Compute ROC Curve and $\text{AUC}(\alpha)$ score; \\
Compute $F1(\alpha)$, $\text{Accuracy}(\alpha)$, and $\text{Precision}(\alpha)$ scores; \\
% \
}
Compute $\text{AUC}(\alpha_{\text{max}})=\max_{\alpha} \text{AUC}(\alpha)$; 

     \Return $\text{F1}(\alpha_{\text{max}})$, $\text{Accuracy}(\alpha_{\text{max}})$, and $\text{Precision}(\alpha_{\text{max}})$
% \end{algorithmic}
\end{algorithm}

\subsection{Effective dimension study}
\label{sub:eff_dim_theory}
In the last part of this work, we will discuss the representation power of \gls{qgan}s and classical equivalents for anomaly detection in \gls{hep}. We consider the effective dimension, initially introduced in Ref.~\cite{effective_dim}, which corresponds to a measure of the expressibility of a neural network.
The effective dimension is an expressibility measure that can be applied to quantum as well as classical models. More specifically, this measure quantifies what fraction of the model space can actually be covered with the available system parameters.
In the case of (q)GANs, the effective dimension measures how the (quantum) generator explores the model space for a given amount of parameters.  A higher effective dimension then indicates a higher capacity of the model.
% It is widely used because it corresponds to a upper bound of the generalization error over unseen data. 

% In our A-qGAN framework, we compare the effective dimension of the classical and quantum generators used for anomaly detection. 

\section{Results and discussion}
\label{ch:results}
In the following, we apply the proposed quantum anomaly detection technique to the \gls{hep} data sets introduced in Section~\ref{sub:dataset}. 
First, we present the results for the training of the (q)\gls{gan}s on up to 8 features, and then also demonstrate its feasibility on actual quantum hardware using 3 features.
It should be noted that experiments at larger scale are currently infeasible due to the noise present in today's quantum devices.
%We first present the results of our anomaly detection scheme on a \gls{pca}-compressed data set using between 3 and 8 principal components. We then show how the A-qGAN performs on a quantum device using 3 compressed features. 
Finally, we study the effective dimension of the applied generators, in order to asses their learning capacity. % of this approach for anomaly detection.

\subsection{Numerical results}
\label{sub:study}
% In this section, we present the results obtained by extracting the first three principal components of the features of the \gls{sm} data set.  

% The classical and quantum anomaly schemes differ in how they are implemented. Once the (quantum) \gls{gan}s have learned the \gls{sm} distribution, the classification of every testing data point is done with an optimization of the anomaly loss function w.r.t. the random seed $z$ in the classical framework, while the quantum case only requires the measurement of quantum circuits. 

%We initially randomly select ten different \gls{sm} subsets of the data set presented in Section~\ref{sub:dataset}. Each data set contains 1'000 data points to train the classical \gls{gan}; from each, we extract 100 data points to train the quantum \gls{gan}. For each anomaly that we try to detect, here, the Graviton and the Higgs boson, we evaluate the model w.r.t. 2 data sets, i.e., \gls{sm} and anomalous events by computing the anomaly score.  

In the quantum simulations, we first optimize the \gls{qgan}s on a noiseless simulator for 500 epochs with the AMSGRAD optimizer~\cite{amsgrad}, applying a learning rate of $10^{-3}$, and $(\beta_{1}, \beta_2)=(0.7, 0.99)$. Once trained, we compute the anomaly score for the \gls{sm} testing, the Higgs, and the Graviton data sets. Based on the anomaly score, we then perform the classification between the \gls{sm} and the anomalies using a varying threshold. Next, we repeat the A-qGAN procedure on a noisy simulator that mimics the behavior of the superconducting IBM Quantum processor \textit{ibmq\_belem}. This optimization is performed with the same optimizer but using a learning rate of $10^{-2}$ to handle the noise of the hardware and ten times fewer training epochs. Once trained, we also compute the anomaly score for all data sets and classify the different events.  

We benchmark the quantum models against classical neural networks consisting of fully-connected layers with different non-linear activation functions %. For the same amount of trainable parameters, as in the quantum case, we compare different kinds of architectures of each layer which consists of using different activation functions 
(sigmoid $\sigma$, ReLU, LeakyReLU with parameters $\alpha=0.2$) or dropout layers of probability $0.25$. The architectures of the classical neural networks are chosen, such that they have a number of trainable parameters comparable to the quantum case.

\begin{figure*}[hptb!]
    \centering
    \includegraphics[width=0.9\textwidth]{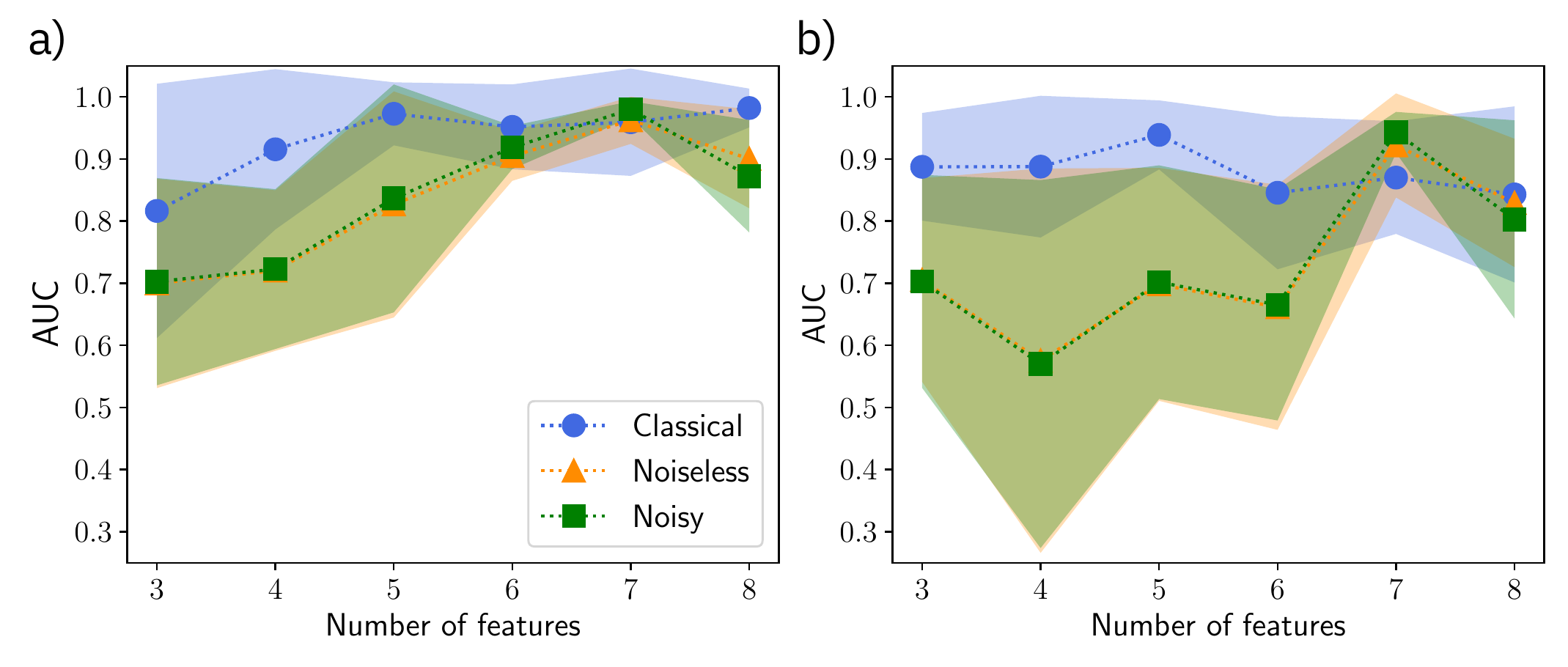}
    \caption{AUC score for increasing number of features used for the detection of the Graviton (a) and the Higgs (b) particles. For the detection of the Graviton, the AUC score increases with the number of features. %until 7 features and decrease for 8 features. 
    For the detection of the Higgs particle, the AUC is constant between 3 and 5 features and slowly decreases after 6 features for the classical model. This is in contrast to the quantum models, which reach their best performance only for 7 features. %However, after 6 features, the quantum simulations get better classification metrics. For both the Graviton and the Higgs particles, the best performance of the quantum model is obtained using 7 features. 
    }
    \label{fig:metrics_evolution}
\end{figure*}

\begin{table*}[htpb!]
    \centering
     \begin{tabular}{c|c|c|c | c|c|c|c}
    \hline 
    \hline
         
         \textbf{Anomaly} & \textbf{Features} & \textbf{Method} &  
         \textbf{Best} $\mathbf{\alpha}$ \textbf{value} &
         \textbf{F1 score} & \textbf{Accuracy} & \textbf{Precision} & \textbf{ROC-AUC}  \\
         \hline
         \multirow{4}{7em}{Graviton} & \multirow{2}{7em}{3} & \textbf{GAN} & 
         $\mathbf{0.25}$ & 
         $\mathbf{0.83 \pm 0.127}$ & $\mathbf{ 0.80 \pm 0.162}$ & $ \mathbf{0.91 \pm 0.183 }$ & $\mathbf{0.82 \pm 0.205}$ \\
         &  & qGAN & $0$ &$ 0.77 \pm 0.084 $& $0.71 \pm 0.126$ & $0.90 \pm 0.163$ & $0.70 \pm 0.169$ \\
          \cline{2-8}
         & \multirow{2}{7em}{7} & \textbf{GAN} & $\mathbf{0.75}$ & $\mathbf{0.96 \pm 0.059} $& $\mathbf{0.95 \pm 0.072} $& $\mathbf{1.0 \pm 0.001}$ & $\mathbf{0.96 \pm 0.086}$ \\
         && qGAN& $0.75$&$0.92 \pm 0.035$ & $0.92 \pm 0.042$ & $1.0 \pm 0.002$&$ 0.96 \pm 0.038$ \\
         \hline 
         \hline
        \multirow{4}{7em}{Higgs} & \multirow{2}{7em}{3} & \textbf{GAN} & $\mathbf{0.5}$ &
         $\mathbf{0.83 \pm 0.088}$ & $\mathbf{0.83 \pm 0.093}$ & $\mathbf{0.98 \pm 0.036} $& $\mathbf{0.89 \pm 0.087}$ \\
         &  & qGAN & $0.25$&$0.75 \pm 0.081 $& $0.70 \pm 0.125$ & $0.87 \pm 0.162$ & $0.71 \pm 0.164$ \\
          \cline{2-8}
         & \multirow{2}{7em}{7} & GAN & $1$& $0.82 \pm 0.089 $& $0.81 \pm 0.098 $& $1.0 \pm 0.003$ & $0.87 \pm 0.091$ \\
         && \textbf{qGAN}& $\mathbf{0.5}$ & $\mathbf{0.89 \pm 0.063}$ & $\mathbf{0.88 \pm 0.080}$ & $\mathbf{1.0 \pm 0.001}$&$ \mathbf{0.92 \pm 0.084}$ \\

         \hline 
         \hline 
         
    \end{tabular}
    \caption{Classification scores for the detection of \gls{bsm} anomalies using three and seven principal components, given for the best $\alpha$ value in the anomaly score. The bold characters correspond to the best results between the classical and quantum methods. In all cases, the quantum calculations are done with 100 data samples while the classical one contains 1000 data samples containing these 100 data samples. For 3 features, the classical model yields better classification scores than for the quantum one for both the Graviton and Higgs detections. For 7 features, the quantum and classical models yield similar classification scores for detection of both the Graviton and Higgs events. %The Higgs detection also gives comparable results for the quantum simulation than the classical one.  
    }
    \label{tab:class_features}
\end{table*}

In the following, we use 10 randomly selected instances of the training and testing data sets introduced in Section~\ref{sub:dataset}.
We perform anomaly detection using between 3 and 8 features, each obtained with \gls{pca}-compression. The classification metrics resulting from the anomaly detection using 3 or 7 features are compared in Table~\ref{tab:class_features}.
The data shows that classical and quantum anomaly detection methods achieve comparable classification metrics in both cases.
For the 3-feature (7-feature) simulation, we use 3 (9) repetitions of the generator ansatz in Fig.~\ref{sfig:generator_ansatz}, and 2 (3) repetitions of the discriminator ansatz in Fig.~\ref{sfig:discriminator_ansatz}, yielding 6 (70) trainable parameters for the generator and 21 (66) trainable parameters for the discriminator, respectively.
%For the 3-feature (respectively for the 7-feature) simulation, the quantum generator is made of 3 repetitions (respectively 9) of single-qubit $R_{Y}$ gates and two-qubit $CZ$ . The quantum discriminator is of the form of Fig.~\ref{sfig:discriminator_ansatz}: it is made of 2 repetitions (respectively 3)  of single-qubit $R_{Y}$, $R_{Z}$ and $R_{Y}$ and two-qubit $CZ$ followed by two-qubit gates $CX$ to the last qubit and 3 single-qubit $R_{X}$, $R_{Y}$ and $R_{Z}$. It yields 6 trainable parameters (respectively 70)  for the generator and 21 (respectively 66) for the discriminator. 
Furthermore, in each training epoch, we apply more optimization updates for the discriminator than for the generator, 5 times more for the training on 3 features, and 10 times more for 7 features, respectively. 
We plot the evolution of the AUC score for an increasing number of features in Fig.~\ref{fig:metrics_evolution}. %which confirms the tendencies previously observed.
% (the other classification metrics are gathered in Appendix~\ref{app:evolution})
As we increase the number of features, the classical model gets better at detecting the Graviton (see Fig.~\ref{fig:metrics_evolution}a). The A-qGAN performance with less training data also improves, reaching the performance of the classical algorithm for six \gls{pca}-compressed features. After 6 features, we reach a plateau where the performance of the classical and quantum models is comparable.
% However, for 8 features, the performance of the quantum algorithm slightly drops and is slightly worse than the classical one. 
Similarly, the classical Higgs detection performance slightly improves as we add more features before reaching a plateau for 6 or more features, where the average AUC scores are worse than those obtained with less features (see Fig.~\ref{fig:metrics_evolution}b). The classical algorithm slightly outperforms the quantum one for less than seven features. For seven and eight features, the quantum simulations with less training data are similar in performance to the classical case. 
Comparing the two figures, we can observe that detecting the Higgs particle is more challenging than detecting the Graviton for both, quantum and classical models, as also noted in Ref.~\cite{julian_hep}. We can also notice that for both, the Graviton and the Higgs detection, we get similar performance between the noiseless and noisy quantum simulations, which indicates some level of noise robustness of the A-qGAN. 
% In particular, at seven features, the noisy A-qGAN performs better for the Higgs detection than the noiseless case, \fta{indicating that the noise introduced during the training helped detect the Higgs particle and thus made the training procedure more robust [This may look weird to a referee, that noise makes things more robust... Overall, we need to be VERY careful: given the standard deviations, the differences in the mean values are not really significant.]}. 

It is worth stressing that the quantum models were trained with 10 times fewer data samples than the classical models, but still achieve comparable accuracy, AUC, precision and F1 scores.
% This gives rise to the hope that in a scaled setting, i.e. more qubits, we could save training resources with the A-qGAN scheme. \OUF{I don't see why this should give rise to the above assumption?}
% It means that we could train our model with fewer resources and give a similar accuracy to the classical method, showing similar conclusions than in Ref.~\cite{fewer_data_qml}. 
%These performance similarities with fewer data could drive future research to optimize the quantum architecture further to find optimal sets of parameters that can outperform the classical algorithm.

%We must also note that based on the plots of Fig.~\ref{fig:metrics_evolution}, the standard deviation for all methods, classical and quantum, is important, making them statistically compatible after 5 features. 
% For the classical detection of the Graviton the standard deviation intervals are wider than the quantum ones, indicating better confidence in the quantum algorithm, especially with 7 features. This result is in contrast with the Higgs detection where the quantum algorithm struggles more to detect the Higgs particle than the Graviton. However, as for the Graviton, the confidence intervals are better for the quantum simulations than for the classical one when using 7 features. 
The large standard deviations shown in Fig.~\ref{fig:metrics_evolution} can be explained by the intrinsic volatility in the \gls{gan} training and by the data set reshufflings.  

\subsection{Hardware experiments} 
\label{sub:hardware}
\begin{figure}[htpb!]
    \centering
    \includegraphics[width=0.45 \textwidth]{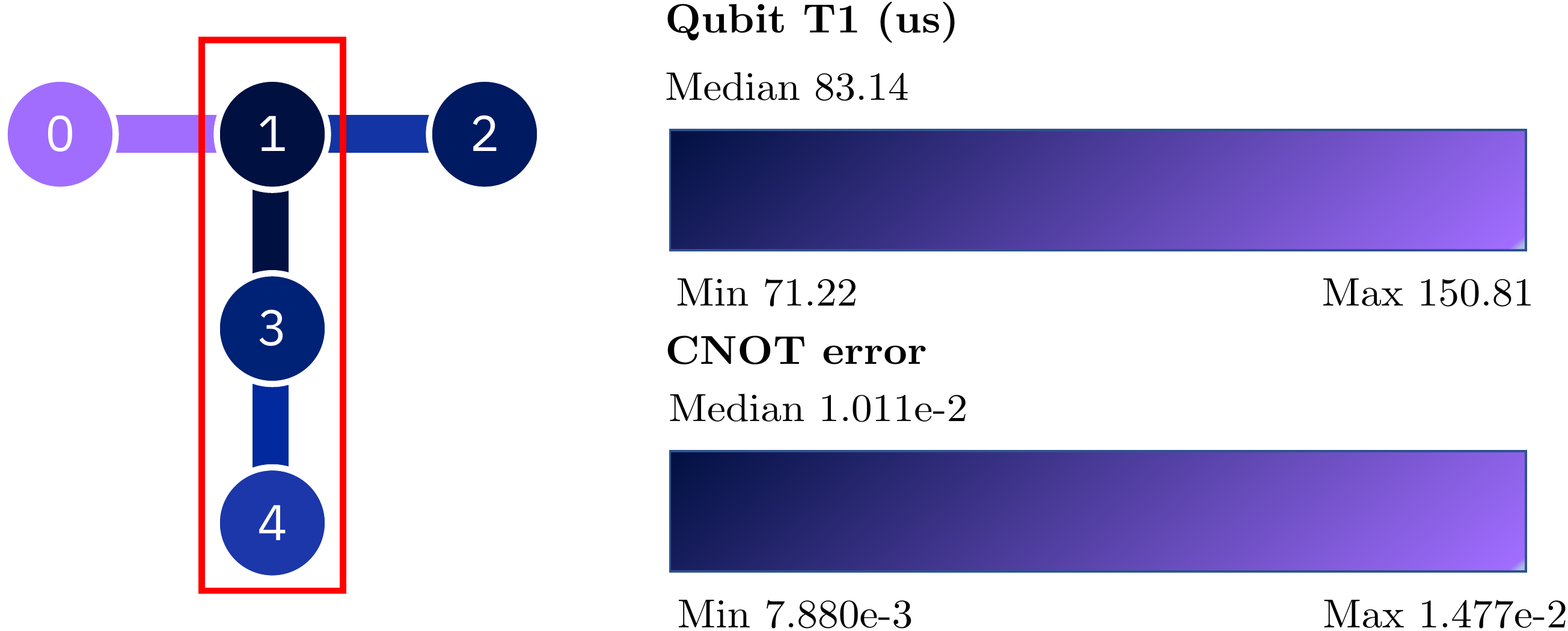}
    \caption{Topology of the IBM Quantum processor \textit{ibmq\_belem}. Qubits are represented as circles and the supported two-qubit gate operations are displayed as edges connecting the qubits. The coloring of the qubits indicates their respective $T_1$ times and the coloring of the edges indicates the errors induced by CNOT gates. The $T_1$ times and CNOT errors were obtained on February 7, 2023. The highlighted qubits are the ones used for the experiments.}
    \label{fig:ibmq_belem}
\end{figure}

Next, we perform anomaly detection on the superconducting IBM Quantum processor \textit{ibmq\_belem}. Fig.~\ref{fig:ibmq_belem} shows the topology of the hardware with the CNOT errors and the $T_1$ time for each qubit. %\fta{T2 would have also been nice, but it's probably too late now...} 
For the training of the \gls{qgan}, we use the same protocol as before (with the same generator and discriminator optimization hyperparameters as in the noisy simulation). We use $10^4$ shots for each expectation value estimation during the training and the anomaly score computation. Additionally, to ease the hardware training, we reduce the number of training epochs to 5.

\begin{figure}[htpb!]
    \centering
    \includegraphics[width=0.4 \textwidth]{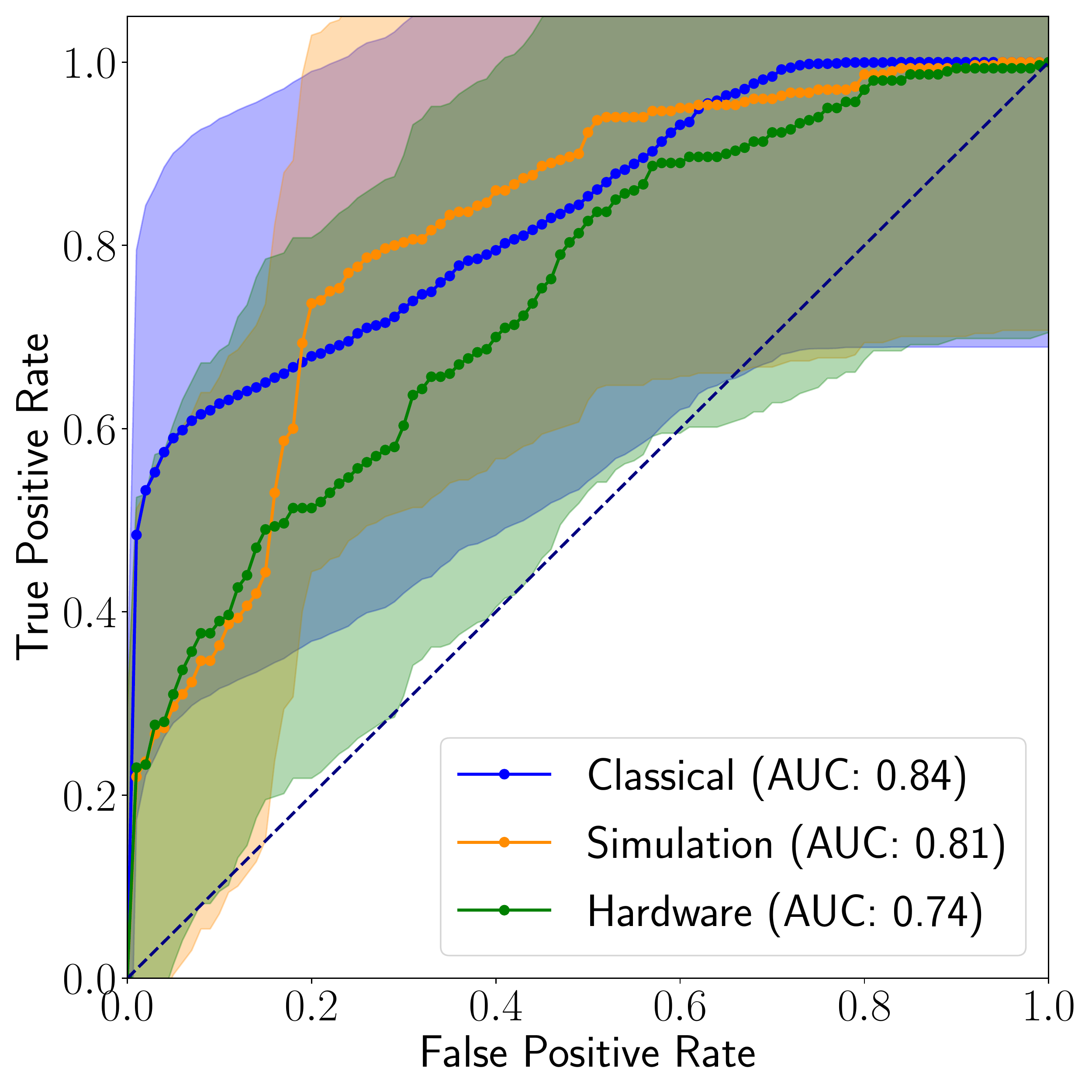}
    \caption{ROC-AUC curve of the classification between \gls{sm} and Graviton events for the classical \gls{gan} (blue), the noiseless simulation of a \gls{qgan} (orange), and the \gls{qgan} executed on the IBM Quantum processor \textit{ibmq\_belem}. All models are trained on 3 features of the \gls{sm} data set, and evaluated on \gls{sm} and Graviton events. %The classical anomaly detection is performed by training the classical \gls{gan} on \gls{sm} and by computing the anomaly score on \gls{sm} and Graviton events (blue).  The quantum approach first consists of training the \gls{qgan} on embedded \gls{sm} events on a noiseless simulator and on the quantum device \textit{ibmq\_belem}. The quantum anomaly detection is performed by computing the anomaly score on the simulator (orange) and on the quantum hardware (green).
    }
    \label{fig:hardware_experiments}
\end{figure}

In this part, we only consider the anomaly detection of the Graviton particle using 3 features. To ease the computation on the quantum hardware, we train on subsets of 50 events taken from the 10 training data sets used in Section~\ref{sub:study} and then compute the anomaly scores for the \gls{sm} and Graviton events. Fig.~\ref{fig:hardware_experiments} presents the average ROC curve of the A-qGAN trained on the noisy quantum hardware (green), compared to the classical method (blue) and the noiseless A-qGAN (orange). 

Compared to the classical and noiseless methods, the AUC curve of the A-qGAN evaluated on the quantum hardware is lower.
These differences can be explained by the reduced number of epochs, and smaller data sets compared to the noiseless algorithm. 

\subsection{Effective dimension study}
\label{sub:eff_dim_result}

We now compare the effective dimensions of the quantum and classical generators for different system sizes. 
It should be noted that  we use a similar amount of trainable parameters for the quantum and classical neural networks employed for both the generator and discriminator of the GAN. Hence, their effective dimensions can be directly compared.
\begin{figure}[htpb!]
    \centering
    \includegraphics[width=.5\textwidth]{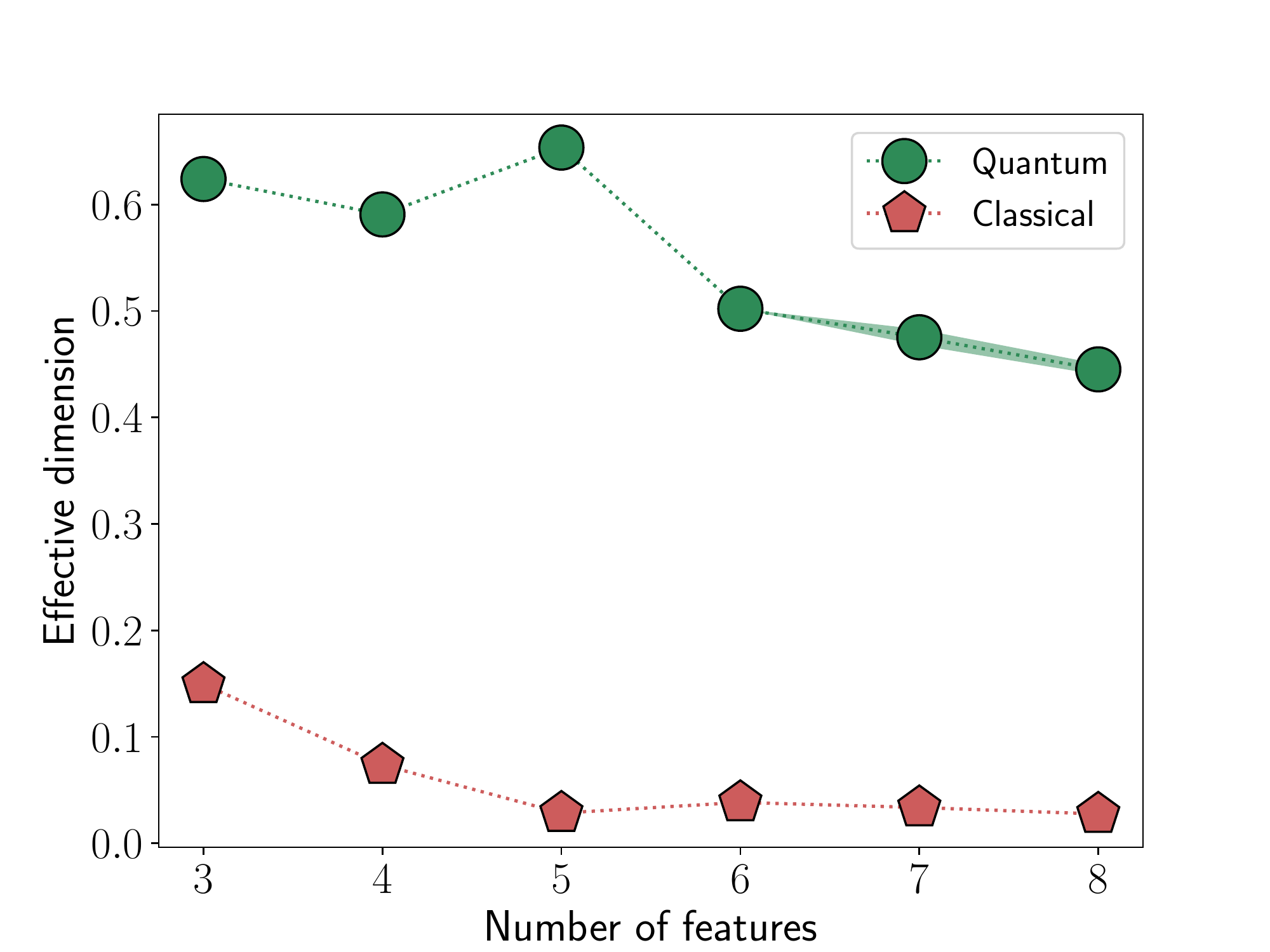}
    \caption{Effective dimension of the classical (red pentagons) and quantum (green circles) generator for increasing number of features. For each number of features, the classical and quantum generator have the same number of trainable parameters. Therefore, the higher effective dimension of the quantum generator indicates a potential advantage of the quantum approach over the classical one.}
    \label{fig:eff_dim}
\end{figure}
Fig.~\ref{fig:eff_dim} shows the effective dimension for increasing number of features. The green (red) dots represent the effective dimension of the trained quantum (classical) model. %as we increase the number of features in our system.
The effective dimension of the quantum model is always larger than for the classical model which indicates a higher expressibility.
It should also be noted that as we increase the dimension of the system, the effective dimension decreases for the chosen generator ansatzes with $\mathcal{O}(n)$ parameters. 
Furthermore, the effective dimension of the classical model appears to converge towards 0, hinting at a lack of capacity when the classical model is restricted to the same amount of trainable parameters as the quantum model. %\fta{What would an effective dimension equal to 0 mean?} \OUF{No capacity} 
This may be indicative for a potential advantage of our quantum methodology in larger dimensions as the effective dimension and, hence, expressive capacity of the classical model for $\mathcal{O}(n)$ parameters could become insufficient.

\section{Discussions and conclusions}
\label{ch:conclusions}
% \fta{[We can probably cut a bit in the `recap' part here, and go more directly to the point, discussing the findings and their implications]}

In this work, we proposed and successfully tested an unsupervised learning strategy based on quantum generative modeling techniques for anomaly detection in HEP. In the introduced A-qGAN algorithm, we apply a \gls{qgan} to learn the underlying probability distribution from a data set of \gls{sm} events encoded in quantum states, and then use it to identify \gls{hep} processes that do not conform with the \gls{sm}. The classification is based on an anomaly score, which serves as a distance measure between \gls{sm} events and other events of interest, not contemplated in the \gls{sm}. Specifically, we verify the proposed anomaly detection scheme by identifying Higgs and Graviton events. %Even though the \gls{hep} data set is obtained by classically post-processing a quantum process, the anomaly detection was tested with a fully quantum approach. The generator is designed to generate quantum states following the \gls{sm} statistics of the encoded data set such that continuous input variables can be supported. 
%A quantum discriminator was chosen to reduce measurement errors and to avoid having a global objective function that may lead to barren plateaus~\cite{bp_1, bp_2}. 
Our quantum anomaly detection scheme is designed to be robust to training instabilities by using an anomaly score that relies on the quantum generator and the quantum discriminator output. Additionally, the proposed generative method can detect new-physics events without modelling a specific \gls{bsm} scenario. 

The presented experiments show that in both cases the anomaly detection performance improves as we increase the number of compressed features with quantum and classical method yielding comparable accuracy. 
Overall, our results indicate that the proposed quantum approach can detect Graviton and Higgs events using ten times fewer data points compared to a classical benchmark based on classical \gls{gan}s with the same number of trainable parameters. 
It should, however, be mentioned that a lack of data is typically not a problem in this type of \gls{hep} task. In fact, the amount of training data will even increase in the next years. Nevertheless, the indication that anomaly detection could be performed with a reduced training data set, could set the ground for other applications suffering from data scarcity in other \gls{hep} tasks or other fields such as medicine. 
% The Graviton detection performance improves as we increase the number of compressed features for both approaches with smaller AUC scores for the quantum algorithm. For detecting the Higgs particle, the quantum algorithm fails to improve upon the classical algorithm (see Fig.~\ref{fig:metrics_evolution}) except in the 7-feature setting. More specifically, with 7 PCA-compressed features, the A-qGAN yields comparable accuracy as the classical benchmark for both anomalies (see Table~\ref{tab:class_features}). 
%
%Despite having smaller classification metrics for each compressed feature 
Moreover, we showed that the quantum model exhibits a larger expressive power, measured by a higher effective dimension%(see Fig.~\ref{fig:eff_dim})
, than the classical counterpart. This suggests that our \gls{qgan}-based anomaly detection scheme could be better suited to model and detect \gls{bsm} anomalies in more complex and/or scaled settings. 

Future research directions aimed at confirming and extending our preliminary results could address a more extensive collection of data sets and scale up the relevant model and feature dimensions, approaching the regime where our A-qGAN can no longer be simulated classically. As the field of unsupervised anomaly detection in \gls{hep} is still relatively young, particularly for what concerns the application of quantum methods, future studies could also investigate alternative approaches, which could be compared or could complement generative models. As an example, it was recently shown that a supervised kernel method based on the generation of a scrambled anomalous data set~\cite{julian_hep} can achieve performances close to the best known classical equivalents, and an unsupervised technique~\cite{cern_qml_new} can potentially outperform its classical counterpart. It is also worth mentioning that the use of classically prepared and pre-processed features might hide some of the quantum correlations originally present in the data~\cite{julian_hep, cern_qml_new}. One could hence envisage the use of the A-qGAN framework for detecting anomalies directly on quantum data sets, obtained for instance through quantum sensing or, e.g., by directly coupling quantum processors to a new generation of quantum detectors.

Lastly, while in this work we mainly focused on detecting anomalies, the proposed \gls{qgan} framework could potentially be modified to facilitate the generation of events beyond the \gls{sm}. In particular, once trained on the \gls{sm}, one could possibly use the model to generate events on a channel orthogonal to the learned data representation. As a result, our framework could open up novel research avenues at the interface between quantum technologies and \gls{hep}.

%In doing so, it could be possible to generate anomalies. However, if the \gls{qgan}s did not learn a proper representation of the \gls{sm}, it is possible to generate - in addition to the desired anomalies -  actual \gls{sm} events as well as unrealistic new-physics events. 
%As for the classical case, the design of quantum algorithms for anomalies detection is still in its infancy and more research is needed before assessing the real potential of this novel technology.     

\section{Acknowledgements}
We acknowledge the use of IBM Quantum services for this work. IBM, the IBM logo, and ibm.com are trademarks of International Business Machines Corp., registered in many jurisdictions worldwide. Other product and service names might be trademarks of IBM or other companies. The current list of IBM trademarks is available at \url{https://www.ibm.com/ legal/copytrade}.
MG and SV are supported by CERN through the Quantum Technology Initiative.

\printglossary
\bibliographystyle{bibliography/IEEEtranN}
\bibliography{bibliography/references_new.bib}

% Generated by IEEEtranN.bst, version: 1.13 (2008/09/30)
\begin{thebibliography}{70}
\providecommand{\natexlab}[1]{#1}
\providecommand{\url}[1]{#1}
\csname url@samestyle\endcsname
\providecommand{\newblock}{\relax}
\providecommand{\bibinfo}[2]{#2}
\providecommand{\BIBentrySTDinterwordspacing}{\spaceskip=0pt\relax}
\providecommand{\BIBentryALTinterwordstretchfactor}{4}
\providecommand{\BIBentryALTinterwordspacing}{\spaceskip=\fontdimen2\font plus
\BIBentryALTinterwordstretchfactor\fontdimen3\font minus
  \fontdimen4\font\relax}
\providecommand{\BIBforeignlanguage}[2]{{%
\expandafter\ifx\csname l@#1\endcsname\relax
\typeout{** WARNING: IEEEtranN.bst: No hyphenation pattern has been}%
\typeout{** loaded for the language `#1'. Using the pattern for}%
\typeout{** the default language instead.}%
\else
\language=\csname l@#1\endcsname
\fi
#2}}
\providecommand{\BIBdecl}{\relax}
\BIBdecl

\bibitem[Grubbs(1969)]{anomaly_1}
F.~E. Grubbs, ``Procedures for detecting outlying observations in samples,''
  \emph{Technometrics}, vol.~11, no.~1, 1969.

\bibitem[Hawkins(1980)]{anomaly_2}
D.~M. Hawkins, \emph{Identification of outliers}.\hskip 1em plus 0.5em minus
  0.4em\relax Springer, 1980, vol.~11.

\bibitem[Rajasegarar et~al.(2008)Rajasegarar, Leckie, and
  Palaniswami]{ano_wireless}
S.~Rajasegarar, C.~Leckie, and M.~Palaniswami, ``Anomaly detection in wireless
  sensor networks,'' \emph{IEEE Wireless Communications}, vol.~15, no.~4, 2008.

\bibitem[Zuech et~al.(2015)Zuech, Khoshgoftaar, and Wald]{ano_intrusion}
R.~Zuech, T.~M. Khoshgoftaar, and R.~Wald, ``Intrusion detection and big
  heterogeneous data: a survey,'' \emph{Journal of Big Data}, vol.~2, no.~1,
  2015.

\bibitem[Quellec et~al.(2016)Quellec, Cazuguel, Cochener, and
  Lamard]{ano_cancer}
G.~Quellec, G.~Cazuguel, B.~Cochener, and M.~Lamard, ``Multiple-instance
  learning for anomaly detection in digital mammography,'' \emph{Ieee
  transactions on medical imaging}, vol.~35, no.~7, 2016.

\bibitem[Apollinari et~al.(2017)Apollinari, B{\'e}jar~Alonso, Br{\"u}ning,
  Fessia, Lamont, Rossi, and Tavian]{hl_lhc}
G.~Apollinari, I.~B{\'e}jar~Alonso, O.~Br{\"u}ning, P.~Fessia, M.~Lamont,
  L.~Rossi, and L.~Tavian, ``High-luminosity large hadron collider. technical
  design report v. 0.1,'' Fermi National Accelerator Lab, Tech. Rep., 2017.

\bibitem[Agliardi et~al.(2022)Agliardi, Grossi, Pellen, and Prati]{qml_mc}
G.~Agliardi, M.~Grossi, M.~Pellen, and E.~Prati, ``Quantum integration of
  elementary particle processes,'' \emph{Physics Letters B}, vol. 832, 2022.

\bibitem[Guan et~al.(2021)Guan, Perdue, Pesah, Schuld, Terashi, Vallecorsa, and
  Vlimant]{ref_qml_hep}
W.~Guan, G.~Perdue, A.~Pesah, M.~Schuld, K.~Terashi, S.~Vallecorsa, and J.-R.
  Vlimant, ``Quantum machine learning in high energy physics,'' \emph{Machine
  Learning: Science and Technology}, vol.~2, no.~1, 2021.

\bibitem[T{\"u}ys{\"u}z et~al.(2021)T{\"u}ys{\"u}z, Rieger, Novotny,
  Demirk{\"o}z, Dobos, Potamianos, Vallecorsa, Vlimant, and Forster]{hep_qml_1}
C.~T{\"u}ys{\"u}z, C.~Rieger, K.~Novotny, B.~Demirk{\"o}z, D.~Dobos,
  K.~Potamianos, S.~Vallecorsa, J.-R. Vlimant, and R.~Forster, ``Hybrid quantum
  classical graph neural networks for particle track reconstruction,''
  \emph{Quantum Machine Intelligence}, vol.~3, 2021.

\bibitem[T{\"u}ys{\"u}z et~al.(2020)T{\"u}ys{\"u}z, Carminati, Demirk{\"o}z,
  Dobos, Fracas, Novotny, Potamianos, Vallecorsa, and Vlimant]{hep_qml_2}
C.~T{\"u}ys{\"u}z, F.~Carminati, B.~Demirk{\"o}z, D.~Dobos, F.~Fracas,
  K.~Novotny, K.~Potamianos, S.~Vallecorsa, and J.-R. Vlimant, ``Particle track
  reconstruction with quantum algorithms,'' in \emph{EPJ Web of Conferences},
  vol. 245.\hskip 1em plus 0.5em minus 0.4em\relax EDP Sciences, 2020.

\bibitem[Terashi et~al.(2021)Terashi, Kaneda, Kishimoto, Saito, Sawada, and
  Tanaka]{hep_qml_3}
K.~Terashi, M.~Kaneda, T.~Kishimoto, M.~Saito, R.~Sawada, and J.~Tanaka,
  ``Event classification with quantum machine learning in high-energy
  physics,'' \emph{Computing and Software for Big Science}, vol.~5, 2021.

\bibitem[{CDF Collaboration} et~al.(2022){CDF Collaboration}, Aaltonen, Amerio,
  Amidei, Anastassov, Annovi, Antos, Apollinari, Appel, Arisawa,
  et~al.]{ref_article_cern}
{CDF Collaboration}, T.~Aaltonen, S.~Amerio, D.~Amidei, A.~Anastassov,
  A.~Annovi, J.~Antos, G.~Apollinari, J.~Appel, T.~Arisawa \emph{et~al.},
  ``High-precision measurement of the {W} boson mass with the {CDF} {II}
  detector,'' \emph{Science}, vol. 376, no. 6589, 2022.

\bibitem[Baldi et~al.(2014)Baldi, Sadowski, and Whiteson]{supervised_1}
P.~Baldi, P.~Sadowski, and D.~Whiteson, ``Searching for exotic particles in
  high-energy physics with deep learning,'' \emph{Nature communications},
  vol.~5, no.~1, 2014.

\bibitem[Chakraborty et~al.(2019)Chakraborty, Lim, and Nojiri]{supervised_2}
A.~Chakraborty, S.~H. Lim, and M.~M. Nojiri, ``Interpretable deep learning for
  two-prong jet classification with jet spectra,'' \emph{Journal of High Energy
  Physics}, vol. 2019, no.~7, 2019.

\bibitem[Ren et~al.(2021)Ren, Wang, Wu, Yang, and Zhang]{supervised_3}
J.~Ren, D.~Wang, L.~Wu, J.~M. Yang, and M.~Zhang, ``Detecting an axion-like
  particle with machine learning at the {LHC},'' \emph{Journal of High Energy
  Physics}, vol. 2021, no.~11, 2021.

\bibitem[Lv et~al.(2022)Lv, Wang, and Wu]{supervised_4}
H.~Lv, D.~Wang, and L.~Wu, ``Deep learning jet images as a probe of light
  {Higgsino} dark matter at the {LHC},'' \emph{Physical Review D}, vol. 106,
  no.~5, 2022.

\bibitem[Nachman(2022)]{model_independent}
B.~Nachman, ``Anomaly detection for physics analysis and less than supervised
  learning,'' in \emph{Artificial Intelligence for High Energy Physics}.\hskip
  1em plus 0.5em minus 0.4em\relax World Scientific, 2022.

\bibitem[Aad et~al.(2020)Aad, Abbott, Abbott, Abud, Abeling, Abhayasinghe,
  Abidi, AbouZeid, Abraham, Abramowicz, et~al.]{bsm_1}
G.~Aad, B.~Abbott, D.~C. Abbott, A.~A. Abud, K.~Abeling, D.~K. Abhayasinghe,
  S.~H. Abidi, O.~AbouZeid, N.~L. Abraham, H.~Abramowicz \emph{et~al.},
  ``{Dijet Resonance Search with Weak Supervision Using $\sqrt{s}= 13$ {TeV} pp
  Collisions in the {ATLAS Detector}},'' \emph{Physical review letters}, vol.
  125, no.~13, 2020.

\bibitem[{CMS Collaboration} et~al.(2021){CMS Collaboration}, Sirunyan,
  Tumasyan, Adam, Ambrogi, Bergauer, Dragicevic, Er{\"o}, Del~Valle,
  Fr{\"u}hwirth, et~al.]{bsm_2}
{CMS Collaboration}, A.~Sirunyan, A.~Tumasyan, W.~Adam, F.~Ambrogi,
  T.~Bergauer, M.~Dragicevic, J.~Er{\"o}, A.~E. Del~Valle, R.~Fr{\"u}hwirth
  \emph{et~al.}, ``{MUSiC: a model-unspecific search for new physics in
  proton--proton collisions at $\sqrt{s}= 13$ TeV},'' \emph{The European
  Physical Journal C}, vol.~81, 2021.

\bibitem[Kasieczka et~al.(2021)Kasieczka, Nachman, Shih, Amram, Andreassen,
  Benkendorfer, Bortolato, Brooijmans, Canelli, Collins, et~al.]{bsm_3}
G.~Kasieczka, B.~Nachman, D.~Shih, O.~Amram, A.~Andreassen, K.~Benkendorfer,
  B.~Bortolato, G.~Brooijmans, F.~Canelli, J.~H. Collins \emph{et~al.}, ``{The
  LHC Olympics 2020 a community challenge for anomaly detection in high energy
  physics},'' \emph{Reports on progress in physics}, vol.~84, no.~12, 2021.

\bibitem[Aarrestad et~al.(2022)Aarrestad, van Beekveld, Bona, Boveia, Caron,
  Davies, De~Simone, Doglioni, Duarte, Farbin, et~al.]{bsm_4}
T.~Aarrestad, M.~van Beekveld, M.~Bona, A.~Boveia, S.~Caron, J.~Davies,
  A.~De~Simone, C.~Doglioni, J.~Duarte, A.~Farbin \emph{et~al.}, ``The dark
  machines anomaly score challenge: benchmark data and model independent event
  classification for the lhc,'' \emph{SciPost Physics}, vol.~12, no.~1, 2022.

\bibitem[Fraser et~al.(2022)Fraser, Homiller, Mishra, Ostdiek, and
  Schwartz]{ref_article_anomaly}
K.~Fraser, S.~Homiller, R.~K. Mishra, B.~Ostdiek, and M.~D. Schwartz,
  ``Challenges for unsupervised anomaly detection in particle physics,''
  \emph{Journal of High Energy Physics}, vol. 2022, no.~3, 2022.

\bibitem[Andreassen et~al.(2019)Andreassen, Feige, Frye, and
  Schwartz]{ano_unsuper_2}
A.~Andreassen, I.~Feige, C.~Frye, and M.~D. Schwartz, ``{JUNIPR}: a framework
  for unsupervised machine learning in particle physics,'' \emph{The European
  Physical Journal C}, vol.~79, 2019.

\bibitem[Dorigo et~al.(2023)Dorigo, Fumanelli, Maccani, Mojsovska, Strong, and
  Scarpa]{ano_unsuper_3}
T.~Dorigo, M.~Fumanelli, C.~Maccani, M.~Mojsovska, G.~C. Strong, and B.~Scarpa,
  ``{RanBox}: anomaly detection in the copula space,'' \emph{Journal of High
  Energy Physics}, vol. 2023, no.~1, 2023.

\bibitem[Schlegl et~al.(2017)Schlegl, Seeb{\"o}ck, Waldstein, Schmidt-Erfurth,
  and Langs]{ref_article3}
T.~Schlegl, P.~Seeb{\"o}ck, S.~M. Waldstein, U.~Schmidt-Erfurth, and G.~Langs,
  ``Unsupervised anomaly detection with generative adversarial networks to
  guide marker discovery,'' in \emph{Information Processing in Medical Imaging:
  25th International Conference}.\hskip 1em plus 0.5em minus 0.4em\relax
  Springer, 2017.

\bibitem[Goodfellow et~al.(2020)Goodfellow, Pouget-Abadie, Mirza, Xu,
  Warde-Farley, Ozair, Courville, and Bengio]{GAN_paper}
I.~Goodfellow, J.~Pouget-Abadie, M.~Mirza, B.~Xu, D.~Warde-Farley, S.~Ozair,
  A.~Courville, and Y.~Bengio, ``Generative adversarial networks,''
  \emph{Communications of the ACM}, vol.~63, no.~11, 2020.

\bibitem[Goodfellow(2016)]{gan_tuto}
I.~Goodfellow, ``{NIPS} 2016 tutorial: {Generative} adversarial networks,''
  \emph{arXiv preprint arXiv:1701.00160}, 2016.

\bibitem[Schlegl et~al.(2019)Schlegl, Seeb{\"o}ck, Waldstein, Langs, and
  Schmidt-Erfurth]{fanogan2}
T.~Schlegl, P.~Seeb{\"o}ck, S.~M. Waldstein, G.~Langs, and U.~Schmidt-Erfurth,
  ``{f-AnoGAN}: Fast unsupervised anomaly detection with generative adversarial
  networks,'' \emph{Medical image analysis}, vol.~54, 2019.

\bibitem[Denton et~al.(2015)Denton, Chintala, Fergus,
  et~al.]{efficient_gan_sample_1}
E.~L. Denton, S.~Chintala, R.~Fergus \emph{et~al.}, ``Deep generative image
  models using a laplacian pyramid of adversarial networks,'' \emph{Advances in
  neural information processing systems}, vol.~28, 2015.

\bibitem[Radford et~al.(2015)Radford, Metz, and
  Chintala]{efficient_gan_sample_2}
A.~Radford, L.~Metz, and S.~Chintala, ``Unsupervised representation learning
  with deep convolutional generative adversarial networks,'' \emph{arXiv
  preprint arXiv:1511.06434}, 2015.

\bibitem[Lloyd and Weedbrook(2018)]{qGAN_1}
S.~Lloyd and C.~Weedbrook, ``Quantum generative adversarial learning,''
  \emph{Physical review letters}, vol. 121, no.~4, 2018.

\bibitem[Dallaire-Demers and Killoran(2018)]{qGAN_3}
P.-L. Dallaire-Demers and N.~Killoran, ``Quantum generative adversarial
  networks,'' \emph{Physical Review A}, vol.~98, no.~1, 2018.

\bibitem[Abbas et~al.(2021)Abbas, Sutter, Zoufal, Lucchi, Figalli, and
  Woerner]{effective_dim}
A.~Abbas, D.~Sutter, C.~Zoufal, A.~Lucchi, A.~Figalli, and S.~Woerner, ``The
  power of quantum neural networks,'' \emph{Nature Computational Science},
  vol.~1, no.~6, 2021.

\bibitem[Berezniuk et~al.(2020)Berezniuk, Figalli, Ghigliazza, and
  Musaelian]{eff_dim_1}
O.~Berezniuk, A.~Figalli, R.~Ghigliazza, and K.~Musaelian, ``A scale-dependent
  notion of effective dimension,'' \emph{arXiv preprint arXiv:2001.10872},
  2020.

\bibitem[Rissanen(1996)]{eff_dim_2}
J.~J. Rissanen, ``Fisher information and stochastic complexity,'' \emph{IEEE
  transactions on information theory}, vol.~42, no.~1, 1996.

\bibitem[Cover(1999)]{eff_dim_3}
T.~M. Cover, \emph{Elements of information theory}.\hskip 1em plus 0.5em minus
  0.4em\relax John Wiley \& Sons, 1999.

\bibitem[Kadurin et~al.(2017)Kadurin, Nikolenko, Khrabrov, Aliper, and
  Zhavoronkov]{druGAN}
A.~Kadurin, S.~Nikolenko, K.~Khrabrov, A.~Aliper, and A.~Zhavoronkov,
  ``{druGAN}: an advanced generative adversarial autoencoder model for de novo
  generation of new molecules with desired molecular properties in silico,''
  \emph{Molecular pharmaceutics}, vol.~14, no.~9, 2017.

\bibitem[Killoran et~al.(2017)Killoran, Lee, Delong, Duvenaud, and
  Frey]{DNAgan}
N.~Killoran, L.~J. Lee, A.~Delong, D.~Duvenaud, and B.~J. Frey, ``Generating
  and designing {DNA} with deep generative models,'' \emph{arXiv preprint
  arXiv:1712.06148}, 2017.

\bibitem[Kurach et~al.(2019)Kurach, Lu{\v{c}}i{\'c}, Zhai, Michalski, and
  Gelly]{gan_origin_2}
K.~Kurach, M.~Lu{\v{c}}i{\'c}, X.~Zhai, M.~Michalski, and S.~Gelly, ``A
  large-scale study on regularization and normalization in {GANs},'' in
  \emph{International conference on machine learning}.\hskip 1em plus 0.5em
  minus 0.4em\relax PMLR, 2019.

\bibitem[Robbins and Monro(1951)]{sgd}
H.~Robbins and S.~Monro, ``A stochastic approximation method,'' \emph{The
  annals of mathematical statistics}, 1951.

\bibitem[Kingma and Ba(2014)]{adam}
D.~P. Kingma and J.~Ba, ``Adam: {A} method for stochastic optimization,''
  \emph{arXiv preprint arXiv:1412.6980}, 2014.

\bibitem[Reddi et~al.(2019)Reddi, Kale, and Kumar]{amsgrad}
S.~J. Reddi, S.~Kale, and S.~Kumar, ``On the convergence of adam and beyond,''
  \emph{arXiv preprint arXiv:1904.09237}, 2019.

\bibitem[Zoufal et~al.(2019)Zoufal, Lucchi, and Woerner]{qGAN_2}
C.~Zoufal, A.~Lucchi, and S.~Woerner, ``Quantum generative adversarial networks
  for learning and loading random distributions,'' \emph{npj Quantum
  Information}, vol.~5, no.~1, 2019.

\bibitem[Kerenidis and Prakash(2020)]{analytic_gradient}
I.~Kerenidis and A.~Prakash, ``Quantum gradient descent for linear systems and
  least squares,'' \emph{Physical Review A}, vol. 101, no.~2, 2020.

\bibitem[Schuld et~al.(2019)Schuld, Bergholm, Gogolin, Izaac, and
  Killoran]{parameter_shift_1}
M.~Schuld, V.~Bergholm, C.~Gogolin, J.~Izaac, and N.~Killoran, ``Evaluating
  analytic gradients on quantum hardware,'' \emph{Physical Review A}, vol.~99,
  no.~3, 2019.

\bibitem[Mitarai et~al.(2018)Mitarai, Negoro, Kitagawa, and
  Fujii]{parameter_shift_2}
K.~Mitarai, M.~Negoro, M.~Kitagawa, and K.~Fujii, ``Quantum circuit learning,''
  \emph{Physical Review A}, vol.~98, no.~3, 2018.

\bibitem[McClean et~al.(2018)McClean, Boixo, Smelyanskiy, Babbush, and
  Neven]{bp_1}
J.~R. McClean, S.~Boixo, V.~N. Smelyanskiy, R.~Babbush, and H.~Neven, ``Barren
  plateaus in quantum neural network training landscapes,'' \emph{Nature
  communications}, vol.~9, no.~1, 2018.

\bibitem[Ortiz~Marrero et~al.(2021)Ortiz~Marrero, Kieferov\'a, and
  Wiebe]{Wiebe2020Barren}
\BIBentryALTinterwordspacing
C.~Ortiz~Marrero, M.~Kieferov\'a, and N.~Wiebe, ``Entanglement-induced barren
  plateaus,'' \emph{PRX Quantum}, vol.~2, Oct 2021.
\BIBentrySTDinterwordspacing

\bibitem[Cerezo et~al.(2021)Cerezo, Sone, Volkoff, Cincio, and Coles]{bp_2}
M.~Cerezo, A.~Sone, T.~Volkoff, L.~Cincio, and P.~J. Coles, ``Cost function
  dependent barren plateaus in shallow parametrized quantum circuits,''
  \emph{Nature communications}, vol.~12, no.~1, 2021.

\bibitem[Thanasilp et~al.(2022)Thanasilp, Wang, Cerezo, and
  Holmes]{fidelity_scable}
S.~Thanasilp, S.~Wang, M.~Cerezo, and Z.~Holmes, ``Exponential concentration
  and untrainability in quantum kernel methods,'' \emph{arXiv preprint
  arXiv:2208.11060}, 2022.

\bibitem[Tacchino et~al.(2021)Tacchino, Mangini, Barkoutsos, Macchiavello,
  Gerace, Tavernelli, and Bajoni]{tacchinoIEEE2021}
F.~Tacchino, S.~Mangini, P.~K. Barkoutsos, C.~Macchiavello, D.~Gerace,
  I.~Tavernelli, and D.~Bajoni, ``Variational learning for quantum artificial
  neural networks,'' \emph{IEEE Transactions on Quantum Engineering}, vol.~2,
  2021.

\bibitem[Knapp et~al.(2021)Knapp, Cerri, Dissertori, Nguyen, Pierini, and
  Vlimant]{top}
O.~Knapp, O.~Cerri, G.~Dissertori, T.~Q. Nguyen, M.~Pierini, and J.-R. Vlimant,
  ``Adversarially {Learned} {Anomaly} {Detection} on {CMS} {Open} {Data}:
  re-discovering the top quark,'' \emph{The European Physical Journal Plus},
  vol. 136, no.~2, 2021.

\bibitem[Buckley et~al.(2011)Buckley, Butterworth, Gieseke, Grellscheid,
  H{\"o}che, Hoeth, Krauss, L{\"o}nnblad, Nurse, Richardson,
  et~al.]{monte_carlo}
A.~Buckley, J.~Butterworth, S.~Gieseke, D.~Grellscheid, S.~H{\"o}che, H.~Hoeth,
  F.~Krauss, L.~L{\"o}nnblad, E.~Nurse, P.~Richardson \emph{et~al.},
  ``General-purpose event generators for {LHC} physics,'' \emph{Physics
  Reports}, vol. 504, no.~5, 2011.

\bibitem[De~Favereau et~al.(2014)De~Favereau, Delaere, Demin, Giammanco,
  Lemaitre, Mertens, and Selvaggi]{monte_carlo_2}
J.~De~Favereau, C.~Delaere, P.~Demin, A.~Giammanco, V.~Lemaitre, A.~Mertens,
  and M.~Selvaggi, ``{DELPHES} 3: a modular framework for fast simulation of a
  generic collider experiment,'' \emph{Journal of High Energy Physics}, vol.
  2014, no.~2, 2014.

\bibitem[Bakhet et~al.(2015)Bakhet, Khlopov, and Hussein]{monte_carlo_graviton}
N.~Bakhet, M.~Y. Khlopov, and T.~Hussein, ``Phenomenology of large extra
  dimensions models at hadrons colliders using monte carlo techniques (spin-2
  graviton),'' \emph{arXiv preprint arXiv:1507.03888}, 2015.

\bibitem[dat({\natexlab{a}})]{dataset_higgs}
CMS collaboration (2016). Simulated dataset
  VBFHiggs0MToBB\_M-125p6\_7TeV-JHUGenV4-pythia6-tauola in AODSIM format for
  2011 collision data (BSM Higgs). CERN Open Data Portal.
  DOI:\href{doi.org/10.7483/OPENDATA.CMS.3R3P.5JYR}{10.7483/OPENDATA.CMS.3R3P.5JYR}.

\bibitem[dat({\natexlab{b}})]{dataset_graviton}
CMS collaboration (2016). Simulated dataset
  Graviton2BPqqbarToZZTo4L\_M-125p6\_7TeV-JHUGenV3-pythia6 in AODSIM format for
  2011 collision data (BSM Higgs). CERN Open Data Portal.
  DOI:\href{doi.org/10.7483/OPENDATA.CMS.SZWT.H9MC}{10.7483/OPENDATA.CMS.SZWT.H9MC}.

\bibitem[Cerri et~al.(2019)Cerri, Nguyen, Pierini, Spiropulu, and Vlimant]{hlf}
O.~Cerri, T.~Q. Nguyen, M.~Pierini, M.~Spiropulu, and J.-R. Vlimant,
  ``Variational autoencoders for new physics mining at the lhc,'' \emph{Journal
  of High Energy Physics}, vol. 2019, no.~5, 2019.

\bibitem[Hotelling(1933)]{pca}
H.~Hotelling, ``Analysis of a complex of statistical variables into principal
  components.'' \emph{Journal of educational psychology}, vol.~24, no.~6, 1933.

\bibitem[Grant et~al.(2018)Grant, Benedetti, Cao, Hallam, Lockhart, Stojevic,
  Green, and Severini]{angle_encoding}
E.~Grant, M.~Benedetti, S.~Cao, A.~Hallam, J.~Lockhart, V.~Stojevic, A.~G.
  Green, and S.~Severini, ``Hierarchical quantum classifiers,'' \emph{npj
  Quantum Information}, vol.~4, no.~1, 2018.

\bibitem[Stoudenmire and Schwab(2016)]{angle_encoding_2}
E.~Stoudenmire and D.~J. Schwab, ``Supervised learning with tensor networks,''
  \emph{Advances in neural information processing systems}, vol.~29, 2016.

\bibitem[Cao et~al.(2020)Cao, Wossnig, Vlastakis, Leek, and
  Grant]{angle_encoding_3}
S.~Cao, L.~Wossnig, B.~Vlastakis, P.~Leek, and E.~Grant, ``Cost-function
  embedding and dataset encoding for machine learning with parametrized quantum
  circuits,'' \emph{Physical Review A}, vol. 101, no.~5, 2020.

\bibitem[{Qiskit contributors}(2023)]{qiskit}
{Qiskit contributors}, ``Qiskit: An open-source framework for quantum
  computing,'' 2023.

\bibitem[McClean et~al.(2016)McClean, Romero, Babbush, and Aspuru-Guzik]{pqc}
J.~R. McClean, J.~Romero, R.~Babbush, and A.~Aspuru-Guzik, ``The theory of
  variational hybrid quantum-classical algorithms,'' \emph{New Journal of
  Physics}, vol.~18, no.~2, 2016.

\bibitem[Benedetti et~al.(2019)Benedetti, Lloyd, Sack, and Fiorentini]{hea}
M.~Benedetti, E.~Lloyd, S.~Sack, and M.~Fiorentini, ``Parameterized quantum
  circuits as machine learning models,'' \emph{Quantum Science and Technology},
  vol.~4, no.~4, 2019.

\bibitem[Nielsen and Chuang(2002)]{qc}
M.~A. Nielsen and I.~Chuang, ``Quantum computation and quantum information,''
  2002.

\bibitem[Hanley and McNeil(1982)]{auc_1}
J.~A. Hanley and B.~J. McNeil, ``The meaning and use of the area under a {ROC}
  curve.'' \emph{Radiology}, vol. 143, no.~1, 1982.

\bibitem[DeLong et~al.(1988)DeLong, DeLong, and Clarke-Pearson]{auc_2}
E.~R. DeLong, D.~M. DeLong, and D.~L. Clarke-Pearson, ``Comparing the areas
  under two or more correlated receiver operating characteristic curves: a
  nonparametric approach,'' \emph{Biometrics}, 1988.

\bibitem[Schuhmacher et~al.(2023)Schuhmacher, Boggia, Belis, Puljak, Grossi,
  Pierini, Vallecorsa, Tacchino, Barkoutsos, and Tavernelli]{julian_hep}
J.~Schuhmacher, L.~Boggia, V.~Belis, E.~Puljak, M.~Grossi, M.~Pierini,
  S.~Vallecorsa, F.~Tacchino, P.~Barkoutsos, and I.~Tavernelli, ``Unravelling
  physics beyond the standard model with classical and quantum anomaly
  detection,'' \emph{arXiv preprint arXiv:2301.10787}, 2023.

\bibitem[Wo{\'z}niak et~al.(2023)Wo{\'z}niak, Belis, Puljak, Barkoutsos,
  Dissertori, Grossi, Pierini, Reiter, Tavernelli, and
  Vallecorsa]{cern_qml_new}
K.~A. Wo{\'z}niak, V.~Belis, E.~Puljak, P.~Barkoutsos, G.~Dissertori,
  M.~Grossi, M.~Pierini, F.~Reiter, I.~Tavernelli, and S.~Vallecorsa, ``Quantum
  anomaly detection in the latent space of proton collision events at the
  {LHC},'' \emph{arXiv preprint arXiv:2301.10780}, 2023.

\end{thebibliography}

\end{document}